\documentclass{aa}

\usepackage{epsfig}
\newcommand{\msun}{\mbox{$M_{\odot}$}}
\newcommand{\Msun}{\mbox{$M_{\odot}$}}
\newcommand{\lsun}{\mbox{$L_{\odot}$}}
\newcommand{\Lsun}{\mbox{$L_{\odot}$}}
\newcommand{\rsun}{\mbox{$R_{\odot}$}}

\newcommand{\teff}{\mbox{$T_{\rm eff}$}}
\newcommand{\Teff}{\mbox{$T_{\rm eff}$}}
\newcommand{\Meff}{\mbox{$M_{\rm eff}$}}
\newcommand{\vinf}{\mbox{$v_{\infty}$}}
\newcommand{\vesc}{\mbox{$v_{\rm esc}$}}
\newcommand{\ratio}{\mbox{v$_{\infty}$/v$_{\rm esc}$}}
\newcommand{\mdot}{\mbox{$\dot{M}$}}
\newcommand{\Mdot}{\mbox{$\dot{M}$}}
\newcommand{\trad}{\mbox{$T_{\rm R}(r,j)$}}
\newcommand{\msunyr}{\mbox{$M_{\odot} {\rm yr}^{-1}$}}

\begin{document}

\thesaurus{07(02.18.7; 08.05.1; 08.13.2; 08.19.3; 08.23.3)}

\title{On the nature of the bi-stability jump in the winds of early-type supergiants}

\author{Jorick S. Vink\inst{1}
 \and Alex de Koter\inst{2,3}
 \and Henny J.G.L.M. Lamers\inst{1}
}
\offprints{Jorick S. Vink, j.s.vink@astro.uu.nl}

\institute{ Astronomical Institute, Utrecht University,
            P.O.Box 80000, NL-3508 TA Utrecht, The Netherlands
            \and
            Astronomical Institute 'Anton Pannekoek', University of Amsterdam,
            Kruislaan 403, NL-1098 SJ Amsterdam, The Netherlands
            \and
            Advanced Computer Concepts, Code 681, Goddard Space Flight Center, 
            Greenbelt, MD 20771, USA 
            }

\titlerunning{The nature of the Bi-stability jump}
\authorrunning{Jorick S. Vink et al.}

\maketitle

\begin{abstract}

We study the origin of the bi-stability jump in the terminal velocity 
of the winds of supergiants near spectral type B1. Observations show
that here the ratio
$\ratio$ drops steeply from about 2.6 at types earlier than B1 to
a value of $\ratio$=1.3 at types later than B2. To this purpose,
we have calculated wind models and 
mass-loss rates for early-type supergiants in a $\teff$ grid covering
the range between $\teff = 12~500$ and 
$40~000~\rm K$. 
These models show the existence of a 
jump in mass loss around $\teff = 25~000~\rm K$ for normal supergiants,
with \mdot\ increasing by about a factor five from $\Teff \simeq 27~500$ 
to 22~500 K for constant luminosity. 
The wind efficiency number $\eta=\mdot \vinf / (L_*/c)$ 
also increases drastically by a factor of 2~-~3
near that temperature. 

We argue that the jump in mass loss is accompanied by a decrease
of the ratio $\ratio$, which is the observed bi-stability jump in 
terminal velocity.
Using self-consistent models for two values of $\teff$, we have 
derived $\ratio$ = 2.4 for \teff\ = 30 000 K and $\ratio$ = 1.2
for \teff\ = 17 500 K. This is within 10 percent of the observed
values around the jump.

Up to now, a theoretical explanation of the observed bi-stability jump
was not yet provided by radiation driven wind theory.
To understand the origin 
of the bi-stability jump, we have investigated the line acceleration for  
models around the jump in detail. 
These models demonstrate that {\it $\mdot$ increases 
around the bi-stability jump due to 
an increase in the line acceleration of Fe~{\sc iii} {\it below} 
the sonic point}. This shows that the mass-loss rate of
B-type supergiants is very sensitive to the abundance and the
ionization balance of iron.

Furthermore, we show that the elements C, N and O are important
line drivers in the {\it supersonic} part of the wind. The 
{\it subsonic} part of the wind is dominated by the line acceleration
due to Fe. Therefore, CNO-processing is expected {\it not} to have a large
impact on \mdot\, but it might have impact on the terminal velocities.

Finally, we discuss the possible role of the bi-stability jump on 
the mass loss during typical variations of Luminous Blue Variable stars.


\end{abstract}


\section{Introduction}
\label{sec:intro}

In this paper we investigate the origin and the consequences 
of the {\it bi-stability jump} of the stellar winds of
early-type stars near spectral type B1. This bi-stability jump
is observed as a steep decrease in the terminal velocity
of the winds from $\vinf \simeq 2.6 \vesc$ for supergiants 
of types earlier than B1 to $\vinf \simeq 1.3 \vesc$ for 
supergiants of types later than B1 (Lamers et al. 1995).
We will show that this jump in the wind velocity is accompanied \
by a jump in the mass-loss rate with
$\Mdot$ increasing by about a factor of five for supergiants
with \Teff\ between 27~500 and 22~500 K.
  
The theory of radiation driven winds predicts that the mass-loss rates
and terminal velocities of the winds of early-type stars
depend smoothly on the stellar parameters, with $\vinf \simeq 3 \vesc$
and $\Mdot \propto L^{1.6}$ (Castor et al. 1975, Abbott 1982, 
Pauldrach et al. 1986, Kudritzki et al. 1989). 
This theory has not yet been applied to predict the observed
jump in the ratio $\ratio$ for supergiants near spectral type B1.
The change 
from a fast to a slow wind is called the $bi$-$stability$ jump. 
If the wind momentum $\Mdot \vinf$ were about constant across
the bi-stability jump, it would imply that the mass-loss rate
would {\it increase} steeply by about a factor of two from stars 
with spectral types earlier than B1 to later than B1.  
Unfortunately, there are no reliable mass-loss determinations from 
observations for stars later than spectral type B1. 

So far, a physical explanation of the nature of this 
bi-stability jump has been lacking. In this paper, we attempt 
to provide such an explanation and we investigate the
change in mass-loss rate that is accompanied by the change in $\vinf$. 

The concept of a bi-stability jump was first described by 
Pauldrach \& Puls (1990) in connection to their model calculations
of the wind of the Luminous Blue Variable (LBV) star 
P~Cygni (\teff = 19.3 kK). Their models showed that small 
perturbations in the basic parameters of this star
can either result in a wind with a relatively high mass loss,
but low terminal velocity, {\em or} in a wind with relatively
low \mdot, but high \vinf. 
Their suggestion was that the mechanism is related to the behaviour 
of the Lyman continuum. If the Lyman continuum exceeds a certain 
optical depth, then as a consequence, the ionization of the 
metals shifts to a lower stage. This causes a larger line 
acceleration $g_{\rm L}$ and finally yields a jump in $\mdot$.
 
The models of Pauldrach \& Puls (1990) for P Cygni show that the wind momentum 
loss per second, $\mdot \vinf$, is about constant on both sides of the jump 
(see Lamers \& Pauldrach 1991). So Lamers et al. (1995) put forward the idea 
that the mass-loss rate for normal stars could increase by about a factor 
of two, if $\vinf$ decreases by a factor of two, so that $\mdot \vinf$ is 
constant on both sides of the jump. 

Whether this is indeed the case, is still unknown. To investigate 
the behaviour of the mass loss at the bi-stability jump, we will derive 
mass-loss rates for a grid of wind models over a range in \teff.
The main goal of the paper is to understand the processes
that cause the bi-stability jump. Although our results are 
based on complex numerical simulations, we have attempted to provide 
a simple picture of the relevant physics. 
We focus on the observed 
bi-stability jump for normal supergiants.
Nevertheless, these results may also provide valuable insight into the possible 
bi-stable winds of LBVs.

It is worth mentioning that 
Lamers \& Pauldrach (1991) and Lamers et al. (1999) 
suggested that the bi-stability 
mechanism may be responsible for the outflowing disks around 
rapidly-rotating B[e] stars. Therefore our results may also
provide information about the formation of rotation induced
bi-stable disks.

The paper is organized in the following way. In Sect. \ref{sec:simple} 
we describe the basic stellar wind theory.
In particular we concentrate on the question: ``what determines \Mdot\ and 
\vinf ?''. We show that \Mdot\ is determined by the radiative acceleration
in the {\it subsonic} region. 
In Sect.~\ref{sec:method} we explain the  
method that we use to calculate the radiative acceleration
with a Monte Carlo technique 
and the mass-loss rates of a grid of stellar parameters.
Sect. \ref{sec:isa} describes the properties 
of the models for which we predict \mdot.
In Sect.~\ref{sec:predictions} 
our predicted bi-stability jump 
in mass loss will be presented. Then, in
Sect.~\ref{sec:origin} we discuss the
origin of this jump
and show that it is due to 
a shift in the ionization balance of Fe {\sc iv} to Fe {\sc iii}.
Then, we discuss the possible role of the 
bi-stability jump in \mdot\ on the variability of LBV stars in 
Sect.~\ref{sec:lbv}. Finally, in Sect.~\ref{sec:concl}, the study will be 
summarized and discussed.


\section{What determines \Mdot\ and \vinf ?}
\label{sec:simple}

\subsection{The theory of \Mdot\ determination}

Mass loss from early-type stars is due to radiation pressure
in lines and in the continuum (mainly by electron scattering).
Since the radiative acceleration by line processes is the dominant
contributor, the winds are ``line-driven'', i.e. the momentum
of the radiation is transferred to the ions by line scattering
or line absorption. Line-scattering and line absorption occur
at all distances in the wind, from the photosphere up to distances of
tens of stellar radii. So the radiative acceleration of the wind
covers a large range in distance.

The equation of motion of a stationary stellar wind is

\begin{equation}
\label{eq:motion}
v\frac{dv}{dr} =- \frac{GM_*}{r^2} - \frac{1}{\rho}\frac{dp}{dr} +
g_{\rm rad}(r) 
\end{equation}
where $g_{\rm rad}$ is the radiative acceleration.
Together with the mass continuity equation

\begin{equation}
\label{eq:continuity}
\Mdot~=~ 4 \pi r^2 \rho(r) v(r)
\end{equation}
and the expression for the gas pressure $p= \cal{R}\rho T/ \mu$,
where $\cal{R}$ is the gas constant and $\mu$ is the mean mass
per free particle in units of $m_H$, we find the equation of motion

\begin{equation}
\label{eq:vdvdr1}
v~\frac{dv}{dr}~=~\left\{\frac{2 a^2}{r} -\frac{G M_{\rm eff}}{r^2}+ g_{\rm L} \right\}~
/~\left\{ 1 - \frac{a^2}{v^2}\right\}
\end{equation}
where $a$ is the isothermal speed of sound. For simplicity we have
assumed that the atmosphere is isothermal.
In this expression the effective mass $M_{\rm eff}=M_*(1-\Gamma_e)$
is corrected for the radiation pressure by electron scattering.
$g_{\rm L}$ is the line acceleration.
The equation has a singularity at 
the point where $v(r)=a$, this 
critical point is the sonic point.
If the line acceleration $g_{\rm L}(r)$ is known as a function of
$r$, the equation can be solved numerically. A 
smoothly accelerating wind solution requires that the numerator of 
Eq. \ref{eq:vdvdr1} reaches zero exactly at the sonic point where the
denominator vanishes. 


It should be stated that {\it this} critical point (sonic point) at 
$r_c \simeq 1.025 R_*$ and $v_c \simeq$ 20 km s$^{-1}$ is {\it not} 
the same as the CAK critical point. The CAK critical point
is located much further out in the wind at $r_c \simeq 1.5 R_*$ 
and about $v_c \simeq 0.5 \vinf$.
If the line acceleration $g_{\rm L}$ in Eq. \ref{eq:vdvdr1} were to be 
rewritten as a function of velocity gradient instead of radius, then one 
would find the CAK critical point.
Pauldrach et al. (1986) showed that if the finite disk correction to the CAK 
theory is applied, then the Modified CAK critical point moves inward and is 
located at $r_c \simeq 1.04 R_*$ and at $v_c \simeq 100$ km s$^{-1}$. 
This is much closer to the sonic point!
Although the (Modified) CAK critical solution may well provide
the correct mass-loss rate and terminal velocity, there is concern about its 
physical reality (see e.g. Lucy 1998 and Lamers \& Cassinelli 1999 for a thorough 
discussion). Lucy (1998) has given arguments favouring the sonic point as the 
physical more meaningful critical point.
We will use the sonic point as the physically relevant critical point. This is the point
where the mass-loss rate is fixed. Throughout the paper
we will therefore refer to the {\it subsonic} part of the wind for the region close to the photosphere
where the mass loss is determined, and to the {\it supersonic} part for the region
beyond the sonic point where the mass-loss rate is already fixed, but the 
velocity has still to be determined.

The critical solution can be found by numerically integrating  
Eq. \ref{eq:vdvdr1},
starting from some lower boundary $r_0$ in the photosphere, 
with pre-specified values of $T_0$ and $\rho_0$
and with a trial value of $v_0$. The value of $v_0$ that produces a 
velocity law 
that passes smoothly through the critical point is the correct value.
Alternatively, for a non-isothermal wind with a pre-specified
$T(\tau)$-relation,
one can integrate inwards from the 
critical point with an assumed location $r_c$, and then adjust this value
until the inward solution gives a density structure that reaches
$\tau=2/3$ at the location where $T(r)=\Teff$ (e.g. see
Pauldrach et al. 1986). The critical solution specifies the values
of $r_0 \simeq R_*$, $\rho_0$ (given by $\tau(r_0)=2/3$) and $v_0$ 
at the lower boundary. This fixes
the value of \mdot\ via the mass continuity 
equation (Eq. \ref{eq:continuity}). Note that \mdot\ is determined 
by the conditions in the {\it subsonic} region!

We will show below 
that an {\it increase} in $g_{\rm L}(r)$ in the subsonic region results
in an {\it increase} in \mdot. This can be understood because in the
{\it subsonic} region, where the denominator of Eq. \ref{eq:vdvdr1} is
negative, an increase in $g_{\rm L}$ gives a smaller velocity gradient.
Integrating from the sonic point inwards to the lower boundary 
with a smaller velocity gradient, implies that the velocity 
the lower boundary should be higher and hence the mass-loss rate,
$\Mdot = 4 \pi r_0^2 \rho_0 v_0$, must be higher. On the other hand, an 
increase in $g_{\rm L}$ in the {\it supersonic} region, yields a larger
velocity gradient and this would directly increase the terminal velocity
\vinf.

Another way to understand how an increase in \Mdot\ is caused by an increase
in $g_{\rm L}$ below the sonic point, is based on the realization that the
density structure of the subsonic region is approximately
that of a static atmosphere. This can be seen 
in Eq. \ref{eq:motion}. Since 
the term $v~dv/dr$ is much smaller than the acceleration of gravity,
it can approximately be set to zero in the
subsonic region. (This is not correct close to the sonic point.)
In an isothermal static atmosphere the density structure follows the
pressure scaleheight. Adding an extra outward force in the
subsonic region results in an increase of the pressure-scaleheight and hence
in a slower outward decrease in density. This means that just below 
the sonic point, where $v \simeq a$, the density $\rho$ will be higher than without the extra force.
Applying the mass continuity equation (Eq.~\ref{eq:continuity}) at the sonic point
then shows that the mass-loss rate will be higher than without
the extra force in the subsonic region.
(See Lamers \& Cassinelli 1999 for a thorough  discussion).


\subsection{A simple numerical experiment: the sensitivity
of \Mdot\ on the subsonic $g_{\rm L}$}
\label{sec:experiment}

\begin{figure}
 \centerline{\psfig{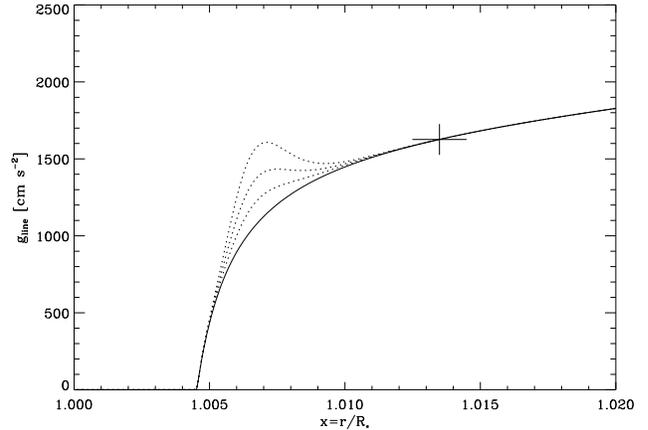}}
 \caption{Extra ``bumps'' on the radiative
          acceleration $g_{\rm L}(r)$ below the sonic point. 
          The solid line is $g_{\rm L}(r)$ of the model without a ``bump''.
          The dotted lines show $g_{\rm L}(r)$ with the adopted bumps  
          with peakheights of 150, 300 and 500 cm s$^{-2}$. 
          The cross indicates the sonic point at 1.0135 $R_*$.}
\label{f_testbumps}
\end{figure}

A simple numerical experiment serves to demonstrate the 
dependence of \mdot\ on the radiative acceleration
in the subsonic region. We start with an isothermal model 
of the wind from a star of 
$M_{\rm eff} = 20 \Msun$, $R_* = 16.92 R_\odot$, $\Teff$ = 25~000 K,
$T_{\rm wind} = 0.8 $\Teff$ = 20~000$ K. We then specify the 
line acceleration $g_{\rm L}(r)$ in such a way that  
it produces a stellar wind with a mass-loss rate of $1.86~10^{-7} \msunyr$
and with a $\beta$-type wind velocity law 

\begin{equation}
\label{eq:betalaw}
v(r)~=~\vinf~(1~-~\frac{R_*}{r})^\beta
\end{equation}
where $\beta$ = 1 and $\vinf = 1500$ km s$^{-1}$. (This $g_{\rm L}(r)$ is found by 
solving Eq. \ref{eq:vdvdr1} with this fixed velocity law).
This model is very similar to one of the models near the
bi-stability jump that we will
calculate in detail in Sect. \ref{sec:predictions}.
As a lower boundary we choose the point where $\rho=10^{-10}$ g cm$^{-3}$
at $r_0=R_*$.
Figure \ref{f_testbumps} shows the resulting variation of $g_{\rm L}(r)$.
Adopting this variation of $g_{\rm L}(r)$ and solving the momentum equation
with the condition that the solution goes smoothly through the sonic 
point, we retrieve the input mass-loss rate and input velocity law, as one would expect.
The sonic point is located at $r_c=1.0135 R_*$ where $v=16.6$ km s$^{-1}$,
and where $g_{\rm L}(r)=1.63~10^3$ cm s$^{-2}$.

Let us study what happens to \Mdot\ and \vinf\ if we change
the line acceleration in the subsonic region. To this purpose
we add a Gaussian ``bump'' to $g_{\rm L}(r)$. This bump is characterized by

\begin{equation}
g_{\rm L}^{\rm bump}(r)~ =~ g_{\rm L}^{\rm peak}~
     \exp \left\{ - \left(\frac{z-z_p}{\Delta z} \right)^2 \right\}
\end{equation}
where $z=1/\{(r/R_*)-1\}$, $z_p=150$ describes the location of the
peak at $r/R_*=1.0067$ and $\Delta z=30$ gives the width of the bump
($\Delta r \simeq 0.0015 R_*$).
The line acceleration with the extra bumps is shown
in Fig. \ref{f_testbumps}.

\begin{figure}
 \centerline{\psfig{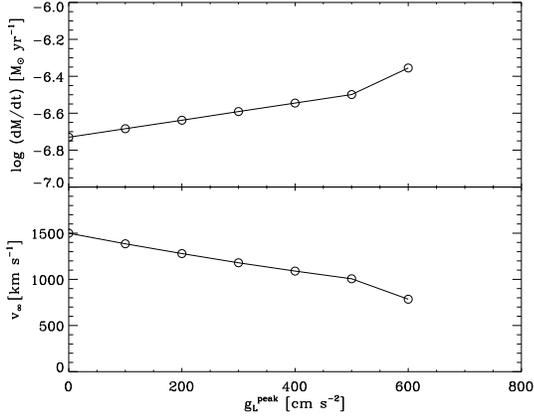}}
 \caption{The effect of increasing the line acceleration in the
          subsonic region on \Mdot\ (upper panel) and 
          a simple derivation
          of its effect on \vinf (lower panel). 
          The horizontal axis
         gives the peak-value, $g_{\rm L}^{\rm peak}$, of the bump in $g_{\rm L}(r)$ 
         in the subsonic region (i.e. the bumps in Fig. 1). }
\label{fig:models}
\end{figure}

The solution of the momentum equation, with the condition that it
passes smoothly through the sonic point, gives the
velocity at the lower boundary and hence the mass-loss rate.
The upper panel of Fig. \ref{fig:models} shows the resulting mass-loss rates
as a function of the peak value of the bump in the line acceleration 
in the subsonic region.
We see that
as the line acceleration in the subsonic region increases,
\mdot\ increases.

\subsection{The effect of an increased \Mdot\ on \vinf}
\label{sec:mdotvinf}

Once \Mdot\ is fixed by the processes in the subsonic region,
the radiative acceleration in the supersonic region then determines the
terminal velocity \vinf\ that the wind will reach. 
This can easily be seen in the following way.
Integrating the momentum equation (Eq. \ref{eq:motion}) in the supersonic
region from the critical point $r_c$ to infinity, and ignoring the influence
of the gas pressure, gives

\begin{equation}
\label{eq:glintegral}
\int_{r_c}^{\infty} g_{\rm L}(r)~dr~=~\frac{1}{2}~\vesc^2~+~\frac{1}{2}~\vinf^2
\end{equation}
so

\begin{equation}
\label{eq:vinfty}
\vinf^2~ \simeq ~2~\int_{r_c}^{\infty} g_{\rm L}(r)~dr~- \vesc^2
\end{equation}
Here we have used the observed property that $\vinf \gg a$ and
that $r_c - r_0 \ll R_*$, so $r_c \simeq R_*$. 
Eq. \ref{eq:vinfty} says that \vinf\ is determined by the 
integral of $g_{\rm L}(r)$ in the {\it supersonic} region.

The radiative acceleration in the {\it supersonic} part of the wind
will {\it decrease} as \mdot\ is forced to increase
by an {\it increase} in the radiative acceleration 
in the {\it subsonic} part of the wind.
This is because the
optical depth of the optically thick driving lines, 
which is proportional to the
density in the wind, will increase. Thus an increase in \Mdot\ results
in an increase of the line optical depth. This results in a decrease
of $g_{\rm L}$ in the supersonic region, which gives a lower
terminal velocity \vinf. We will estimate this effect below.

Assume that the radiative acceleration by lines depends on the
optical depth in the wind, as given by CAK theory
(Castor et al. 1975). 

\begin{equation}
g_{\rm L}(r)~=~g_{\rm e}~M(t)~=~\frac{\sigma_e L_*}{4 \pi r^2 c} k t^{-\alpha}
\label{eq:CAK}
\end{equation}
where $k$ and $\alpha$ are constants and $g_{\rm e}$ is a reference 
value describing the acceleration due to electron scattering. It is given 
by $g_{\rm e}=\frac{\sigma_e L_*}{4 \pi r^2 c}$.
The optical depth parameter is 

\begin{equation}
t~=~ \sigma_e v_{\rm th} \rho (dr/dv)
\label{eq:tCAK}
\end{equation}
where $v_{th}$ is the mean thermal velocity of the protons.
Let us define $g_{\rm L}^{\rm init}(r)$ as the radiative acceleration in the
{\it supersonic} part of the initial wind model, i.e. without
the increased mass-loss rate due to the bump in the subsonic region,
and $g_{\rm L}(r)$ as the radiative acceleration of the model 
with the increased \Mdot.
From Eqs. \ref{eq:CAK} and \ref{eq:tCAK} with Eq. \ref{eq:continuity}
we find that 

\begin{equation}
g_{\rm L}(r)=g_{\rm L}^{\rm init} 
    \left\{ \frac{r^2vdv/dr}{(r^2vdv/dr)^{\rm init}}\right\}^{\alpha}
    \left\{ \frac{\Mdot^{\rm init}}{\Mdot}\right\}^{\alpha}
\label{eq:glratio}
\end{equation}
where the superscript ``init'' refers to the initial model.

Let us now compare the terminal velocities of the initial
model without the bump, to that with the
increased mass-loss rate due to the bump, in a 
simple but crude way,
by solving the momentum equation in the supersonic part of the 
wind. If we neglect the terms due to the gas pressure
and due to the gravity, the momentum equation in the supersonic part
of the wind reduces to

\begin{equation}
v \frac{dv}{dr}~\simeq~g_{\rm L}(r)
\label{eq:simplemoment}
\end{equation}
Solving the equation for the initial model and the model with the increased \mdot\ 
results in the following expression

\begin{equation}
v\frac{dv}{dr}~\simeq~ \left\{ v\frac{dv}{dr}\right\}^{\rm init} 
      \left\{ \frac{\Mdot^{\rm init}}{\Mdot} \right\}^{\alpha/(\alpha-1)}
\label{eq:vdvdr}
\end{equation}
So the ratio between the terminal velocities of the models with 
and without the increased mass-loss rate is

\begin{equation}
\frac {\vinf}{v_{\infty}^{\rm init}}~ \simeq~ 
   \left\{ \frac{\Mdot^{\rm init}}{\Mdot} \right\}^{\alpha/(2-2\alpha)}
     ~\simeq~
   \left\{ \frac{\Mdot}{\Mdot^{\rm init}} \right\}^{-3/4}
\label{eq:vinfratio}
\end{equation}
where we adopted $\alpha = 0.60$ (Pauldrach et al. 1986)
for the last expression.
We see that \vinf\ will decrease roughly as
$\Mdot^{-3/4}$ when the mass-loss rate increases.
The result is shown in the lower panel of Fig. \ref{fig:models}.

We realize that this numerical test is a drastic simplification
of the real situation: 
(a) we have assumed an isothermal wind;
(b) we have taken the lower boundary at a fixed density; 
(c) we have ignored possible changes in the ionization of the
wind due to changes in \Mdot\ and
(d) we have ignored the role of the gas pressure and of gravity in estimating
the change in \vinf.
However, this simple test
serves the purpose of explaining {\it qualitatively} that the mass-loss rate
depends on the radiative acceleration in the subsonic part of the wind only,
and that an increase in the mass-loss rate due to an increase of $g_{\rm L}$
in the subsonic region will also be accompanied by a
decrease in \vinf. In the rest of the paper, we will {\it quantitatively}
calculate radiative accelerations and mass-loss rates with a method which will 
be described in Sect. \ref{sec:method}.

Thus, an {\it increase} in the radiative acceleration 
in the {\it subsonic} region of the wind results in an {\it increase} of 
\mdot\ and a {\it decrease} in \vinf.  
So, in order to understand the origin of the
bi-stability jump of radiation driven winds,
and to predict its effect on \mdot\ and \vinf,
we should pay close attention to the calculated 
radiative acceleration in the {\it subsonic} part of the wind.


\section{The method to predict \mdot}
\label{sec:method}

In order to understand the nature of the bi-stability jump, we calculate
a series of radiation driven wind models for supergiants in the range of
\Teff = 12~500 to 40~000 K. 
The calculation of the radiative acceleration of the winds 
requires the computation of the contributions of a very large number
of spectral lines. To this end, we first calculate the thermal, density and
ionization structure of a wind model computed with the non-LTE expanding
atmosphere code {\sc isa-wind} (de Koter et al. 1993)(for details, see 
Sect. \ref{sec:isa}). We then calculate the
radiative acceleration by following the fate of a very large number 
of photons that are released from below the photosphere into the wind, by means of a 
Monte Carlo technique. In this section, 
we describe the basic physical properties of the 
adopted Monte Carlo (MC) technique which was first applied to
the study of winds of early-type stars by Abbott \& Lucy (1985). 
Then, we describe the calculation of the radiative acceleration
by lines with the MC method, and finally  
the method for calculating theoretical mass-loss 
rates.


\subsection{Momentum transfer by line scattering}
\label{sec:gline}

The lines in the MC method are described in the Sobolev approximation.
This approximation for the line acceleration 
is valid if the physical conditions over a Sobolev length do not 
change significantly, i.e. 

\begin{equation}
\frac{1}{f} \mid{\frac{d f}{ dr}}\mid \ll \frac{1}{v_{\rm t}} \mid{\frac{dv}{dr}}\mid
\label{eq:sob}
\end{equation}
where {\it f} ~is any physically relevant variable for the line driving, e.g.
density, temperature or ionization fraction.
$v_{\rm t}$ is a combination of thermal and turbulent velocities.
Eq. \ref{eq:sob} shows that the validity range of the Sobolev approximation 
is in practice somewhat arbitrary, since it depends on the value of the 
turbulent velocity which is poorly known. Nevertheless, the Sobolev approximation 
is often used (e.g. Abbott \& Lucy 1985) and we will also adopt it 
in calculating the line acceleration and mass loss,
mainly because of computational limitations. We cannot exclude
that due to the use of the Sobolev approximation we may predict quantitatively 
inaccurate line accelerations below the sonic point. 
However, {\it if} an exact treatment 
would be followed, then this is expected to 
have a systematic effect on the line acceleration for {\it all} models. Therefore,
we do not expect our conclusions regarding the origin of the bi-stability jump to be affected.

The Sobolev approximation
implies that 
for scatterings in the frame co-moving with the ions in the wind 
(co-moving frame, CMF), the incident and emerging frequencies are 
both equal to the rest frequency of the line transition $\nu_{0}$ in the CMF.

\begin{equation}
\nu_{\rm in}^{\prime}~=~\nu_{0}~=~\nu_{\rm out}^{\prime}
\end{equation}
where $\nu_{\rm in}^{\prime}$ and $\nu_{\rm out}^{\prime}$ are the 
incident and emerging frequencies in the CMF. In terms of quantities seen 
by an outside observer, these two CMF frequencies are given by:

\begin{equation}
\nu_{\rm in}^{\prime}~=~\nu_{\rm in}~(1~-~\frac{\mu_{\rm in} v}{c})
\label{e_nuin}
\end{equation}
and

\begin{equation}
\nu_{\rm out}^{\prime}~=~\nu_{\rm out}~(1~-~\frac{\mu_{\rm out} v}{c})
\label{e_nuout}
\end{equation}
where $\nu_{\rm in}$ and $\nu_{\rm out}$ are the incident and emergent 
frequencies for an outside observer; $\mu_{\rm in}$ and $\mu_{\rm out}$ 
are the direction cosines with respect to the radial flow velocity of 
the photons at the scattering point and $v$ is the radial flow velocity 
of the scattering ion for an outside observer. Thermal motions of the 
scattering ions are assumed to be negligible compared to the motion of 
the outward flow. Note that we adopted the same velocity $v$ for the ion
before and after the photon interaction (Eqs.~\ref{e_nuin} and 
\ref{e_nuout}). This is justified since the change in velocity due to the 
transfer of momentum from a photon to an ion is very small, i.e. about
$10^1$ cm s$^{-1}$ per scattering.. Therefore, the change in $\nu_{\rm in}$ and
$\nu_{\rm out}$ is mainly determined by a change in direction angle.

Combining Eqs.~\ref{e_nuin} and \ref{e_nuout} gives the conservation of 
co-moving frequency in a scattering event (Abbott \& Lucy 1985).

\begin{equation}
\nu_{\rm in}~(~1~-~\frac{\mu_{\rm in}~v}{c})~=~\nu_{\rm out}~(~1~-~\frac{\mu_{\rm out}~v}{c})
\end{equation}
Because the energy and momentum of a photon are $E = h \nu$ and 
$p = h \nu /c$, the equation can be rewritten in the following way:

\begin{equation}
p_{\rm in}~\mu_{\rm in}~-~p_{\rm out}~\mu_{\rm out}~=~\frac{E_{\rm in}~-E_{\rm out}}{v}
\label{e_pE}
\end{equation}
Eq.~\ref{e_pE} links the change in radial $momentum$ of a photon in an
interaction with an ion with velocity $v$ to the $energy$ loss
of the photon. In order to determine the line acceleration $g_{\rm L}$ 
we will need to derive the momentum transfer from the
photons to the {\it ions} in the wind. 

For an outside observer, the conservation of radial momentum is:

\begin{equation}
m v_{1}~+~\frac{h \nu_{\rm in}}{c}~\mu_{\rm in}~=~m v_{2}~+~\frac{h \nu_{\rm out}}{c}~\mu_{\rm out}
\end{equation}
where $m$ is the mass of the moving ion and $v_{1}$ and $v_{2}$ are the 
radial velocities of the ion just before and after the scattering. For an 
outside observer:

\begin{equation}
\nu_{\rm in}~=~\nu_{0}~(1~+~\frac{\mu_{\rm in} v}{c})
\end{equation}
and

\begin{equation}
\nu_{\rm out}~=~\nu_{0}~(1~+~\frac{\mu_{\rm out} v}{c})
\end{equation}
Again, the change in frequency is dominated by the change in 
direction angle. 
So the change in radial velocity
per scattering, \(\Delta v~=~v_{2} - v_{1} \), is small 
compared to $v$ and is given by

\begin{eqnarray}
\Delta v~&=&~v_{2} - v_{1}  \nonumber \\
         &=&~\frac{h \nu_{0}}{m c} {(1~+~\frac{\mu_{\rm in} v}{c})\mu_{\rm in}~-~\frac{h \nu_{0}}{m c}(1~+~\frac{\mu_{\rm out} v}{c})\mu_{\rm out}}
\label{eq:deltavlarge}
\end{eqnarray}
Since $v \ll c$, Eq. \ref{eq:deltavlarge} becomes

\begin{equation}
\Delta v~=~v_{2} - v_{1}~=~\frac{h \nu_{0}}{m c} (\mu_{\rm in}~-~\mu_{\rm out})
\label{e_deltav}
\end{equation}
This relation describes the velocity increase of the ion depending on the 
directions $\mu_{\rm in}$ and $\mu_{\rm out}$ of the photon. 
In case 
$\mu_{\rm in}$ = $\mu_{\rm out}$ then $\Delta v$ = 0, as one would 
expect. 
The increase in the radial momentum 
\(\Delta p~=~m~\Delta v \) of the scattering ion is now given by:

\begin{eqnarray}
\Delta p~=~m (v_{2} - v_{1})~&=&~\frac{h \nu_{0}}{c} (\mu_{\rm in}~-~\mu_{\rm out})~
 \nonumber \\
  &=&~\frac{h \nu_{\rm in}~-~h \nu_{\rm out}}{v}~=~\frac{\Delta E}{v}
\label{e_gp}
\end{eqnarray}
where \(\Delta E = E_{\rm in} - E_{\rm out} \) is the loss of 
{\it radiative} energy. 
This equation shows that the increase in the momentum of the {\it ions}
can be calculated from the loss of energy of the {\it photons}
when these are followed in their path through the wind by means of the
Monte Carlo method.

%
Multiplying both sides of Eq.~\ref{e_gp} by $v$ and using the fact that
for each scattering 
$v_2 \simeq v_1$ so $v \simeq (v_{1} + v_{2})/2$, gives:

\begin{equation}
\frac{1}{2} m (v_{2}^2~-~v_{1}^2)~=~h \nu_{\rm in}~-~h \nu_{\rm out}
\label{e_energy}
\end{equation}
Equation \ref{e_energy} says that the gain of kinetic energy of the ions in 
the radial direction equals the energy loss of the photons.


\subsection{Scattering and absorption processes in the MC calculations}

The radiative acceleration as a function of distance 
is calculated by means of the 
MC technique by following the fate of the photons 
using the program {\sc mc-wind} (de Koter et al. 1997).
In the calculation of the path of the photons
we have properly taken into account the possibility that the photons 
can be scattered or absorbed \& re-emitted due to true absorption or eliminated 
because they are scattered back into the star.

The radiative transfer in {\sc mc-wind} is calculated in the Sobolev 
approximation. Multiple line and continuum processes are included in 
the code. The continuum processes included are electron scattering and 
thermal absorption and emission. The line processes included are photon 
scattering and photon destruction by collisional de-excitation. 
In deciding whether a continuum or a line event takes place, we have
improved the code in the following way: 
The key point of the Monte-Carlo ``game'' is that {\it line} 
processes can only occur at specific points in each shell of the stellar wind,
whereas {\it continuum} processes can occur at any point. The correct way of 
treating the line and continuum
processes is by comparing a random optical depth 
value to the {\it combined} optical depth for line and continuum processes
along the photon's path. 
First, this combined optical depth is compared to a random
number to decide whether a continuum or a line process takes place.
This first part of the treatment is basically the same as described by 
Mazzali \& Lucy (1993) for the
case of line and electron scatterings only.
Now, after it has been decided that the process will be a continuum process,
a second random number is drawn to decide {\it which} continuum process will
take place, an electron scattering or absorption.


\subsection{The calculation of the radiative acceleration $g_{\rm L}(r)$}
\label{sec:acceleration}

The radiative acceleration of the wind was calculated by following
the fate of the photons emitted from below the photosphere with the
MC technique.
To this purpose the atmosphere is divided into
a large number of concentric, thin shells 
with radius $r$, thickness $\Delta r$ containing a mass $\Delta m(r)$. 

The loss of photon energy due to all scatterings that occur within each 
shell are calculated
to retrieve the total line acceleration $g_{\rm L}(r)$ per shell. The total 
line acceleration per shell summed over all line scatterings in that shell equals

\begin{equation}
g_{\rm L}(r)  = \frac{1}{\Delta m(r)} \frac{\Sigma~\Delta p(r)}{\Delta t}
\end{equation}
where $p(r)$ is the momentum of the ions in the shell.
The momentum gained by the ions in the shell is equal to the momentum
lost by the photons due to interactions in that shell.
Using the relationship between $\Delta m(r)$ and $\Delta r$ for thin 
concentric shells, $\Delta m(r) = 4 \pi r^{2} \rho(r) \Delta r$, and 
the derived relation between momentum and energy transfer of the photons 
$\Delta p = \Delta E /v $ (Eq.~\ref{e_gp}), $g_{\rm L}(r)$ can be 
rewritten as

\begin{equation}
g_{\rm L}(r) = \frac{1}{4 \pi r^{2} \rho(r) \Delta r} \frac{\Sigma~\Delta E(r)}{v(r) \Delta t}
\end{equation}
where $\Sigma \Delta E(r)$ is sum of the energy loss of all the photons 
that are scattered in the shell.
Now using mass continuity (Eq. \ref{eq:continuity})
and the fact that the total energy transfer $\Sigma~\Delta E(r)$ divided 
by the time interval $\Delta t$ equals the rate at which the radiation 
field loses energy, $- \Delta L(r)$, i.e. 
$\Sigma~\Delta E(r) / \Delta t = - \Delta L(r)$, the expression 
for $g_{\rm L}(r)$, which is valid for each shell, simply becomes 
(Abbott \& Lucy 1985)

\begin{equation}
\label{eq:gl}
g_{\rm L} (r)  = -~\frac{1}{\mdot}~\frac{\Delta L(r)}{\Delta r}
\end{equation}
The line list that is used for the MC calculations consists of 
over $10^5$ of the strongest lines of the elements H~-~Zn from 
a line list constructed by 
Kurucz(1988). Lines in the wavelength region between 50 and 7000 \AA~are 
included in the calculations with ionization stages up to stage {\sc vi}.
Typically about $2~10^5$ photon packets, distributed over the 
spectrum at the lower boundary of the atmosphere were followed
for each model, i.e. for each adopted
set of stellar and wind parameters. 
For several more detailed models we calculated the fate of 
$2~10^7$ photon packets. The wind was divided in 
about 50-60 concentric shells, with many narrow shells in the subsonic 
region and wider shells in supersonic layers. 
The division in shells is essentially made on the basis of a Rosseland optical
depth scale. Typical changes in the logarithm of this optical depth are about
0.13.


\subsection{The determination of \mdot}
\label{sec:determination}

We predict the mass-loss rates for a grid of
model atmospheres to study the behaviour of \mdot\
near the bi-stability jump.
For a given set of stellar parameters we calculate the mass loss
in the following way:

\begin{enumerate}

\item{} For fixed values of $L$, $\Teff$, $R_*$ and \Meff\ we
adopt several values of the input mass loss  
$\Mdot^{\rm inp}$ (within reasonable
bounds predicted by CAK theory).

\item{} For each model we adopt a wind with a terminal velocity of
1.3, 2.0 or 2.6 times the effective escape velocity, given by
\begin{equation}
\vesc~=~\sqrt \frac{2 G M_{\rm eff}}{R_*}
\end{equation}
A $\beta$-type velocity law with $\beta=1$ was adopted, appropriate for
OB stars (Groenewegen \& Lamers 1989; Puls et al. 1996)

\item{} For each set of stellar and wind parameters
we calculate a model atmosphere with {\sc isa-wind} (see Sect. \ref{sec:isa}).
This code gives the thermal structure,
the ionization and excitation structure and the population of the
energy levels of all relevant ions.

\item{} For each model the radiative acceleration was calculated
with the {\sc mc-wind} program that uses the Monte Carlo method 
described above.  

\item{} For each set of stellar parameters and 
for each adopted value of $\vinf$ 
we check which one of
the adopted mass-loss rates is consistent with the
radiative acceleration. This consistency was checked in the following way:

Neglecting the term due to the gas pressure, one can write the equation of motion in 
the following way:

\begin{equation}
\label{eq:eqmotion} 
v~\frac{dv}{dr}~=~- \frac{G M_{\rm eff}}{r^2}~+~g_{\rm L}(r)
\end{equation}
 Using the expression for the line acceleration (Eq. \ref{eq:gl}) and 
integrating 
the equation of motion (Eq. \ref{eq:eqmotion}) 
from the stellar surface to infinity gives

\begin{equation}
\label{eq:consistency}
\frac{1}{2}~\mdot~({\vinf}^2~+{\vesc}^2)~=~\Delta L~=~ 
            \mdot \int_{R_*}^\infty g_{\rm L}(r) dr
\end{equation}
$\Delta L = \Sigma \Delta L(r)$, is the total amount of 
radiative energy, summed over all the shells, that 
is lost in the process of line-interaction and is
transfered into kinetic energy of the ions as given in
Eq. \ref{e_energy}. Equation \ref{eq:consistency} states that the momentum
transfered from the radiation into the wind is used to
lift the mass loss out of the potential well and to accelerate
the wind to \vinf.
Only {\it one} value
of \mdot\ will satisfy this equation (Lucy \& Abbott 1993). This 
is the predicted mass-loss rate.
  
\end{enumerate}

We note that Eq. \ref{eq:consistency} only describes the``global''
consistency of the mass-loss rate with the radiative acceleration.
For the set-up of the model atmosphere the velocity law
$v(r)$ is needed as input. This means that although the \Mdot\ calculation
is globally consistent in terms of kinetic wind energy, the velocity
is not necessarily locally consistent, since the equation of motion is not solved.
Instead, we have used observed values for \vinf\ and $\beta$ for the velocity law.
Since the total amount of radiative energy in Eq. \ref{eq:consistency}
is mainly determined in the {\it supersonic} region, where the Sobolev approximation
is an excellent approximation, $\Delta L$ is accurately calculated. 
This implies that if one adopts the correct values for the terminal velocity, 
one may predict accurate values for \mdot !


\section{The model atmospheres}
\label{sec:isa}

The calculation of the mass-loss rates by the method described in the previous 
section requires the input of a model atmosphere, before the
radiative acceleration and \mdot\ can be calculated.

The model atmospheres used for this study are calculated
with the most recent version of the non-LTE unified Improved
Sobolev Approximation code {\sc isa-wind} for stars with
extended atmospheres. For a detailed description of this code we 
refer to de Koter et al. (1993, 1997). 
Here, we just make a few relevant remarks. 

{\sc isa-wind} treats the atmosphere in a unified manner, 
i.e. no artificial separation between photosphere and wind is 
assumed. This is distinct from the so-called ``core-halo'' approaches.
In the photosphere the density structure follows from a 
solution of the momentum equation taking into account gas 
and radiative pressure on electrons. The velocity law
follows from this density structure via the mass continuity equation.
Near the sonic point, a smooth transition is made to a $\beta$-type 
velocity law for the supersonic part of the wind (see Eq. \ref{eq:betalaw}).

The temperature structure in the wind is computed under 
the assumption of radiative equilibrium in an extended grey LTE 
atmosphere. The temperature in the wind is not allowed to drop 
below a certain minimum value $T_{\rm min} = 1/2~\teff$ (Drew 1989).
Finally, the chemical species included explicitly in the non-LTE 
calculations
are H, He, C, N, O and Si. The complexity of the model atoms
is similar to that used by de Koter et al. (1997).
For the iron-group elements, which are important for the
radiative acceleration, we calculate
the ionization/excitation equilibrium in the
{\em modified nebular approximation} (see Schmutz 1991). In this
representation the ionization equilibrium is given by

\begin{equation}
\label{eq:nebular}
   \frac{N_{j+1} n_{\rm e}}{N_{j}} = \{ (1-\zeta)W+\zeta \}W 
      \left( \frac{T_{\rm e}}{T_{\rm R}} \right)^{1/2}
      \left( \frac{N_{j+1} n_{\rm e}}{N_{j}} \right)^{\rm LTE}_{T_{\rm R}}
\end{equation}
where $n_{\rm e}$ and $T_{\rm e}$ are the electron density
and temperature, $N_{j}$ and $N_{j+1}$ are the ion population
numbers, $T_{\rm R}$ = \trad\ is the radiation temperature 
of ion $j$ at radial depth $r$, and $W$ is a geometrical dilution factor 
as defined by Schmutz et al. (1990). 
The last factor of Eq. \ref{eq:nebular} is the LTE ionization
ratio for a temperature \trad.
The parameter $\zeta$, introduced
by Abbott \& Lucy (1985), represents the fraction of recombinations
going directly to the ground state.
The values of \trad\ are obtained by inverting
the above equation, using all 19 ionization ratios available
from the {\sc isa-wind} calculation. 
The
radiation temperature of an explicit ion is used that has its
ionization potential closest (but lower) to that of the metal ion
of interest. For instance, the N~{\sc ii}/{\sc iii} ratio is used to
define the ionization equilibrium of Fe~{\sc iii}/{\sc iv}.  

The excitation state of metastable levels is assumed to be in 
LTE relative to the ground state. For all other levels we adopt
``diluted'' LTE populations, defined by

\begin{equation}
   \frac{n_{u}}{n_{1}} = W \left( 
      \frac{n_{u}}{n_{1}} \right)^{\rm LTE}_{T_{\rm R}} .
\end{equation}
where $n_{u}$ and $n_{l}$ are the excitation population numbers 
for the upper and lower levels.
Clearly, the simplified treatment of the iron-group metals is 
prompted by the computationally intensive nature of the problem
at hand. It needs to be improved in the future, but we do not expect
that our conclusions regarding the nature of the bi-stability
jump would be affected. (We return to this in the discussion in Sect. \ref{sec:concl}).


\section{The predicted bi-stability jump}
\label{sec:predictions}

Using the procedure as described in Sect. \ref{sec:determination}, we 
calculated mass-loss rates for stars with a
luminosity of $L_*=10^5$ \Lsun\ and a mass of $M_*=20 \Msun$.
The models have effective temperatures between 12~500 and 40~000 K
with a stepsize of 2500 K.
These parameters are approximately those of OB supergiants, for which
Lamers et al. (1995) found the bi-stability in \vinf.

We calculated \mdot\ for wind models with a $\beta$-type
velocity law with $\beta = 1$ (Eq. \ref{eq:betalaw}) for three
values of the $\ratio$ = 2.6, 2.0 and 1.3.
Lamers et al. (1995) found that $\ratio \simeq 2.6$ for stars of types
earlier than B1, and $\ratio \simeq 1.3$ for stars of types later than B2. 
For the determination of $\vesc$ we used the effective 
mass $M_{\rm eff} = 17.4~\msun$, with $\Gamma_{e} = 0.130$.

The stellar parameters for the calculated grid are indicated in 
Table \ref{t:parameters}. The models are calculated for solar metallicities.

\subsection{The predicted bi-stability jump in \Mdot}

The results are listed in Table \ref{t:parameters}. This Table
gives the values of \Teff, $R_*$, \vesc\ and  \Mdot\ for each
temperature and for the three values of $\ratio$. We also give the
value of the wind efficiency factor $\eta$, which describes the
fraction of the {\it momentum} of the radiation that is transferred
to the ions

\begin{equation}
\label{eq:eta}
\Mdot \vinf~=~\eta \left( \frac{L_*}{c} \right) 
\end{equation}
The fraction of the photon {\it energy} that is transferred into kinetic
energy of the ions is also listed (in column 8). The values for this
$energy$ efficiency number \(\Delta L/L\) are a factor of about 
$10^{-3}$ smaller than the wind {\it momentum} efficiency 
number $\eta$, which is given in column (7). This is because a photon 
transfers a large fraction of its momentum during a scattering, but only a very 
small fraction (of order $v/c$) of its energy.
The last column of Table \ref{t:parameters} marks three models that will
be discussed in more detail in Sect. \ref{sec:origin}.

\begin{table*}
 \caption[]{Stellar parameters of the grid of calculated models. \\
  log $L/\lsun = 5.0$, $M = 20 \msun$, 
  $\Gamma_{e} = 0.130$, $M_{\rm eff} = 17.4~\msun$,  \( \beta = 1 \), solar metallicity.}
 \label{t:parameters}
  \begin{tabular}{ccrrrcccc}
  \hline
  \noalign{\smallskip}
  $\ratio$ & $\teff$ & $R_{*}$ & $\vesc$ & $\vinf$ & log~\mdot\ & $\eta$ & $\Delta L/L$ & model \\
  \noalign{\smallskip}
        & (K)   & ($\rsun)$ & (km s$^{-1}$) & (km s$^{-1}$) & ($\msun$/yr) & &  (in 10$^{-3}$) & \\
  \hline

   1.3 & 12 500 & 67.7 &  310 &  410 & - 6.32 & 0.095 & 0.103 & \\
       & 15 000 & 47.0 &  380 &  490 & - 6.39 & 0.097 & 0.126 & \\
       & 17 500 & 34.5 &  440 &  570 & - 6.28 & 0.146 & 0.221 & \\
       & 20 000 & 26.4 &  500 &  650 & - 6.22 & 0.192 & 0.332 & \\
       & 22 500 & 20.9 &  560 &  730 & - 6.15 & 0.254 & 0.493 & \\
       & 25 000 & 16.9 &  630 &  810 & - 6.12 & 0.302 & 0.653 & C \\
       & 27 500 & 14.0 &  690 &  900 & - 6.40 & 0.174 & 0.414 & \\
       & 30 000 & 11.8 &  750 &  980 & - 6.58 & 0.126 & 0.326 & \\
       & 32 500 & 10.0 &  810 & 1060 & - 6.58 & 0.136 & 0.382 & \\
       & 35 000 &  8.6 &  880 & 1140 & - 6.43 & 0.207 & 0.626 & \\
       & 37 500 &  7.6 &  940 & 1220 & - 6.37 & 0.255 & 0.826 & \\
       & 40 000 &  6.6 & 1000 & 1300 & - 6.26 & 0.350 & 1.210 & \\

    \smallskip

   2.0 & 12 500 & 67.7 &  310 &  630 & - 6.74 & 0.056 & 0.073 & \\
       & 15 000 & 47.0 &  380 &  750 & - 6.62 & 0.088 & 0.138 & \\
       & 17 500 & 34.5 &  440 &  880 & - 6.49 & 0.139 & 0.254 & \\
       & 20 000 & 26.4 &  500 & 1000 & - 6.41 & 0.191 & 0.398 & \\
       & 22 500 & 20.9 &  560 & 1130 & - 6.32 & 0.264 & 0.620 & \\
       & 25 000 & 16.9 &  630 & 1250 & - 6.48 & 0.203 & 0.530 & \\
       & 27 500 & 14.0 &  690 & 1380 & - 6.73 & 0.125 & 0.360 & \\
       & 30 000 & 11.8 &  750 & 1500 & - 6.76 & 0.128 & 0.400 & \\
       & 32 500 & 10.0 &  810 & 1630 & - 6.71 & 0.155 & 0.527 & \\
       & 35 000 &  8.6 &  880 & 1750 & - 6.59 & 0.220 & 0.801 & \\
       & 37 500 &  7.6 &  940 & 1880 & - 6.57 & 0.247 & 0.969 & \\
       & 40 000 &  6.6 & 1000 & 2000 & - 6.48 & 0.325 & 1.356 & \\

    \smallskip

   2.6 & 12 500 & 67.7 &  310 &  810 & - 6.95 & 0.045 & 0.070 & \\
       & 15 000 & 47.0 &  380 &  980 & - 6.85 & 0.067 & 0.126 & \\
       & 17 500 & 34.5 &  440 & 1140 & - 6.69 & 0.114 & 0.248 & \\
       & 20 000 & 26.4 &  500 & 1300 & - 6.54 & 0.184 & 0.458 & \\
       & 22 500 & 20.9 &  560 & 1460 & - 6.59 & 0.184 & 0.517 & \\
       & 25 000 & 16.9 &  630 & 1630 & - 6.79 & 0.129 & 0.403 & B \\
       & 27 500 & 14.0 &  690 & 1790 & - 6.95 & 0.098 & 0.337 & A \\
       & 30 000 & 11.8 &  750 & 1950 & - 6.92 & 0.115 & 0.430 & \\
       & 32 500 & 10.0 &  810 & 2120 & - 6.86 & 0.143 & 0.579 & \\
       & 35 000 &  8.6 &  880 & 2280 & - 6.76 & 0.194 & 0.845 & \\
       & 37 500 &  7.6 &  940 & 2440 & - 6.71 & 0.233 & 1.089 & \\
       & 40 000 &  6.6 & 1000 & 2600 & - 6.68 & 0.266 & 1.327 & \\

  \noalign{\smallskip}
 \hline

  \end{tabular}
\end{table*}

\begin{figure*}
\centerline{\psfig{file=H1511.f3, width = 12.0cm}}
\centerline{\psfig{file=H1511.f4, width = 12.0cm}}
 \caption{Upper panel: 
  The calculated mass-loss rates \mdot\ as a function of $\teff$ 
  for three values of the ratio 
  $\ratio$. The values for $\ratio$ are indicated in the lower left corner. 
  The stellar parameters are 
  log \( L/\lsun = 5.0 \), $ M = 20~\msun$ and $\beta = 1.0$; all models 
  are calculated for solar metallicities.
  Lower panel:
  The predicted bi-stability jump in $\mdot$ from models with
  the observed ratios of $\ratio = 2.6$ for
  $\teff > 21,000~\rm K$ and $\ratio = 1.3$ for 
  $\teff < 21,000~\rm K$, as indicated in the lower 
  left corner.}
\label{fig:bistability}
\end{figure*}

The results are plotted in Fig.~\ref{fig:bistability}.
For each of the three values of $\ratio$ the value of  
\mdot\ is decreasing for decreasing  $\teff$ between 40~000 and
30~000 K and also between 22~500 and 12~500 K.  
Between about $\teff = 27~500$ K 
and $\teff = 20~000$ K (slightly dependent on $\ratio$)
the mass loss {\it increases} with decreasing \Teff.
These increments in \mdot\ roughly coincide 
in $\teff$ with the observed bi-stability jump in $\ratio$ near 
spectral type B1, at about 21~000 K. 
For the ratio of $\ratio$ = 2.6, the increase
in \mdot\ between model A and B equals 45 \%. We know from the observations 
that $\ratio$ jumps from 2.6 at the hot side of 21~000 K to
1.3 at the cool side of 21~000 K (Lamers et al. 1995).
Including this observed 
jump in $\ratio$ in the mass-loss predictions, provides an even steeper 
increase in \mdot\ from models A and B to the smaller value of $\ratio$ = 1.3,
as is shown in the lower part of Fig.~\ref{fig:bistability}. This figure 
shows an increase in \mdot\ of about a factor of five between  
$\teff = 27~500$ and 20~000 K.  This is our prediction for a bi-stability 
jump in $\mdot$. 

The exact position of $\teff$ of the 
bi-stability jump in Fig.~\ref{fig:bistability} is somewhat ambiguous, 
since $\vinf$ is adopted from observations, and does not directly 
follow from our models. 
For a discussion on the exact 
position of the jump in $\teff$, see Sect.~\ref{sec:concl}.


To test the sensitivity of our predictions of mass-loss rates for different
shapes of the velocity law, we calculated another series of models 
with $\beta$ = 1.5 . Since the differences are only about 10 \%, we conclude
that the predicted mass-loss rates are only marginally sensitive to the 
{\it shape} of the adopted velocity law. 



\subsection{The predicted bi-stability jump in $\eta$}
\label{sec:eta}

\begin{figure*}
 \centerline{\psfig{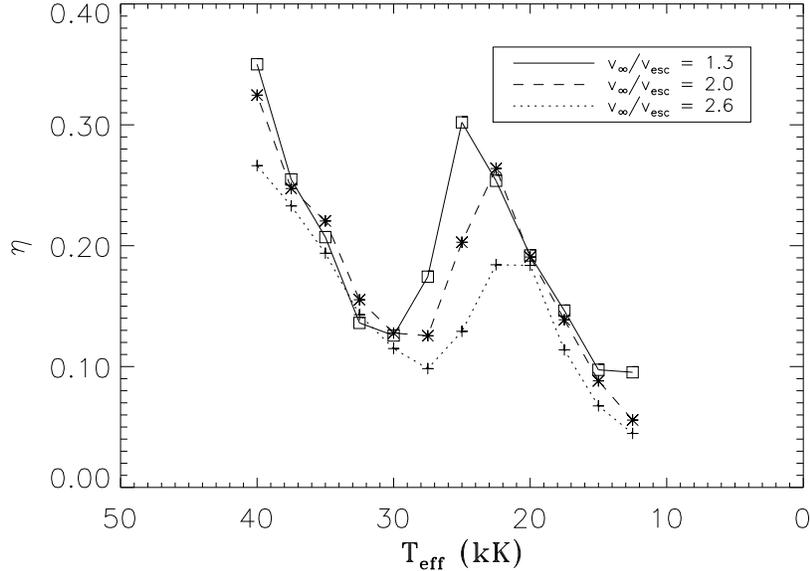}}
 \caption{The wind efficiency number $\eta=\Mdot \vinf  / (L_*/c)$ as a function of $\teff$ for three values of the ratio 
  $\ratio$. These values are indicated in the upper right corner.
  Note the steady decrease of $\eta$ to lower temperatures,
  except the jump of about a factor 2 or 3 near 25~000 K. } 
\label{f_eta}
\end{figure*}

Another view at these results can be obtained by plotting the wind 
efficiency factor $\eta$. Figure \ref{f_eta} shows the behaviour of 
$\eta$ as a function of $\teff$ for the same grid of models as was 
presented for the mass-loss rates in the upper panel of Fig.~\ref{fig:bistability}.

Fig.~\ref{f_eta} clearly shows that $\eta$ is not a constant 
function of $\teff$. The overall picture shows that for the three 
values of $\ratio$, $\eta$ decreases as $\teff$ decreases. This is probably 
due to the fact that the maximum of the flux distribution shifts to 
longer wavelengths. At $\lambda >$ 1800 \AA\ there are significantly less 
lines than at $\lambda <$ 1800 \AA. Therefore, radiative acceleration
becomes less effective at lower $\teff$. In the ranges of 
$40~000 < \teff < 30~000$  and $20~000 < \teff < 12~500$ K, 
$\eta$ is almost independent of the adopted value for $\ratio$. This
means that the behaviour of $\eta$ is intrinsically present in the model 
calculations and does not depend on the values adopted for $\ratio$.

In the range of $30~000 < \teff < 20~000$ K, the 
situation is reversed. $\eta$ now {\it increases} by a 
factor of 2 to 3. This means that the wind momentum loss, 
$\mdot \vinf$ is {\it not} constant over the jump, but 
instead, jumps by a factor of 2~-~3 also. Since $\vinf$ drops by a factor 
of about two, \mdot\ is expected to jump by a factor of about five, which was already
shown in the lower panel of Fig.~\ref{fig:bistability}. 

The behaviour of $\eta$ as a function of \teff\ is not exactly the same 
for the three different series of models. First, the size of the jump
is different. Second, the jump occurs at somewhat different  
temperatures. This is no surprise, since the ionization equilibrium 
does not only depend on $T$, but on $\rho$ as 
well, a smaller value of the velocity $\vinf$, means a larger density $\rho$ 
in the wind. Hence, the jump is expected to start at a larger value 
of $\teff$ for a smaller value of $\ratio$. This behaviour for the 
position of $\teff$ of the jump can be seen in \mdot\ in 
Fig.~\ref{fig:bistability} and in $\eta$ in Fig.~\ref{f_eta}.


\section{The origin of the bi-stability jump}
\label{sec:origin}

In the previous section we have shown
that the mass-loss rate increases around 
$\teff = 25~000$ K. The next step is to investigate the physical process that 
causes the bi-stability jump. Therefore, we will look into 
the details of the line acceleration $g_{\rm L}(r)$ 
for three models around the 
bi-stability jump. For these models (A, B and C in 
Table \ref{t:parameters} and Fig.~\ref{fig:bistability}) we made improved 
Monte-Carlo calculations, using $2 \times 10^7$ packets of photons,
to derive more details about the radiative acceleration.

First, we will investigate the line acceleration $g_{\rm L}(r)$ of the model 
at the hot side of the bi-stability jump. This model A with  
$\teff = 27~500$ K and $\ratio$ = 2.6, is our basic model. 
Then, we will compare model A to model B that has the same 
$\ratio$, but is situated on the cool side of the bistability 
jump, where $\teff$ = 25~000 K. By comparing models A and B, we can 
investigate the intrinsic increase in \mdot\ of 45 \% in our model 
calculations due to the lower $\teff$. 
The next step is to compare $g_{\rm L}(r)$ of model B and model C which also has 
$\teff$ = 25~000 K, but a smaller ratio $\ratio = 1.3$. 
By comparing model B and C, we can obtain 
information about the effects of a jump in $\vinf$.
Finally, we check our approach for self-consistency by
simultaneously calculating the mass-loss rate {\it and} terminal velocity.


\subsection{The main contributors to the line acceleration}
\label{sec:contribute}

\begin{figure*}
 \centerline{\psfig{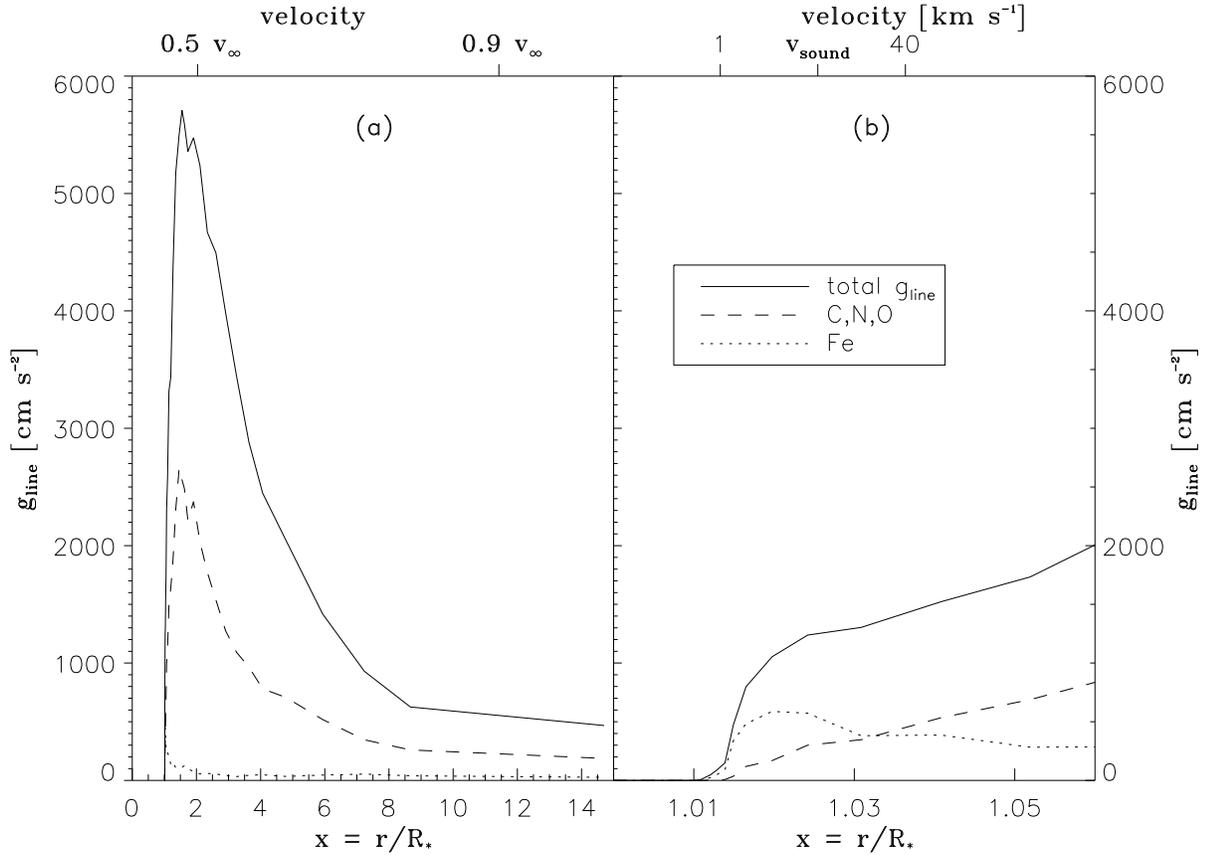}}
 \caption{ The line acceleration of model A (\Teff = 27~500 K and $\ratio=2.6$),
          from 1 to 15 $R_*$ (left) and around the sonic point (right).
          {\bf (a)}~The solid line shows the total $g_{\rm L}$ as a 
          function of the distance.
          The dashed line is the contribution by  
          C, N and O only. The dotted line 
          shows the contribution by Fe lines. 
          Some values
          for the velocity are indicated on top of the figure. 
          {\bf (b)}~ The region around the sonic point is 
          enlarged. The sonic point is reached at x = 1.025.
          Note the bump in the $g_{\rm L}(r)$ just below the sonic point,
          which is largely due to Fe lines.}
\label{f_modelA}
\end{figure*}

Model A has a mass-loss rate of log \mdot\ = -6.95. The behaviour 
of the line acceleration as a function of the distance from the stellar 
surface, $g_{\rm L}(r)$ is shown in Fig.~\ref{f_modelA}.
The sonic point is reached at a distance of 1.025 $R_*$.
It is clear that most of the line driving is produced far beyond 
the sonic point. But, as was explained in Sect.~\ref{sec:simple} the 
important region that determines the mass-loss rate 
is $below$ the sonic point. 
Therefore, the part of the atmosphere around the sonic point is enlarged
in Fig.~\ref{f_modelA}(b).

To investigate the origin of the jump, it is useful to know which elements 
are effective line drivers in which part of the stellar wind. Therefore, 
extra Monte-Carlo calculations were performed. 
The first extra Monte-Carlo simulation was 
performed with a line list containing only Fe lines. The second one was 
performed with a line list containing the lines of the elements C, N and O.

Figure \ref{f_modelA}(b) shows that Fe is the main line driver below the 
sonic point. C, N and O, are important line drivers in the supersonic part 
of the wind, which can be seen in \ref{f_modelA}(a). C, N and O contribute
roughly 50 \% of the line acceleration in the supersonic part of the wind. 
Not indicated here, but relevant to mention is that Si, Cl, P and S are 
other important line drivers in the supersonic part of the wind. Ni was 
found $not$ to be an important line driver in any part of the stellar 
wind at all.

The mass-loss rate is determined by the radiative acceleration
{\it below} the sonic point, and the terminal velocity is determined
by the radiative acceleration in the {\it supersonic} part of the wind.
So our results show that the mass-loss rates of hot star winds
are mainly determined by the radiation pressure due to Fe!
The terminal velocities are mainly determined by the contributions
of C, N and O.


\subsection{The effect of the Fe ionization}
\label{sec:ioniz}

\begin{figure*}
 \centerline{\psfig{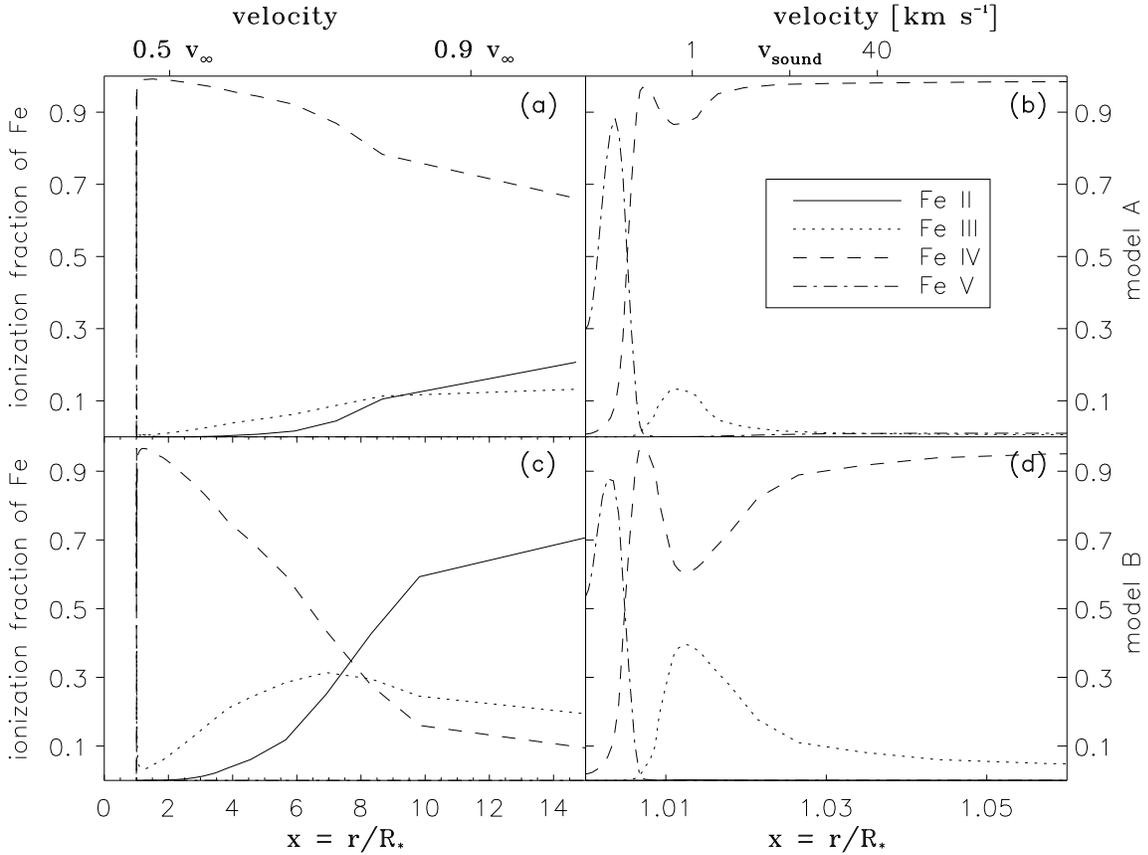}}
 \caption{The ionization fraction of Fe as a function of distance.
          The upper panels are for model A and the lower panels for model B.
          ~{\bf (a)}~Fe ionization for model A from $x=1$ to 15.
          ~{\bf (b)}~Model A, enlarged around sonic point.
          ~{\bf (c)}~Fe ionization for model B from $x=1$ to 15.
          ~{\bf (d)}~Model B, enlarged around sonic point.}
\label{f_ionization}
\end{figure*}

To understand the origin of the bi-stability jump in $\mdot$, we 
investigate the line acceleration due to Fe. The ionization 
balance of Fe for models A and B is plotted in 
Fig.~\ref{f_ionization}, top and bottom respectively.
The right hand figures show the enlargement of the
ionization balance in the region near the sonic point.
In Model A (\Teff=27~500 K) 
Fe V has a maximum around $x = 1.004$, which can be seen in Fig.~\ref{f_ionization} (b). 
Then, due to the outward decreasing temperature, 
Fe V decreases in favour of 
Fe~{\sc iv}, which peaks around $x = 1.008$. 
Next, one may expect Fe~{\sc iv} to decrease in 
favour of Fe~{\sc iii}. However, around $x = 1.013$ 
Fe~{\sc iv} re-ionizes due to a 
decrease of the density $\rho$. In this region of the atmosphere, where 
$dv/dr$ is rapidly increasing, the effect of the decreasing $\rho$ 
is larger than the effect of the decreasing $T$.

\begin{figure*}
 \centerline{\psfig{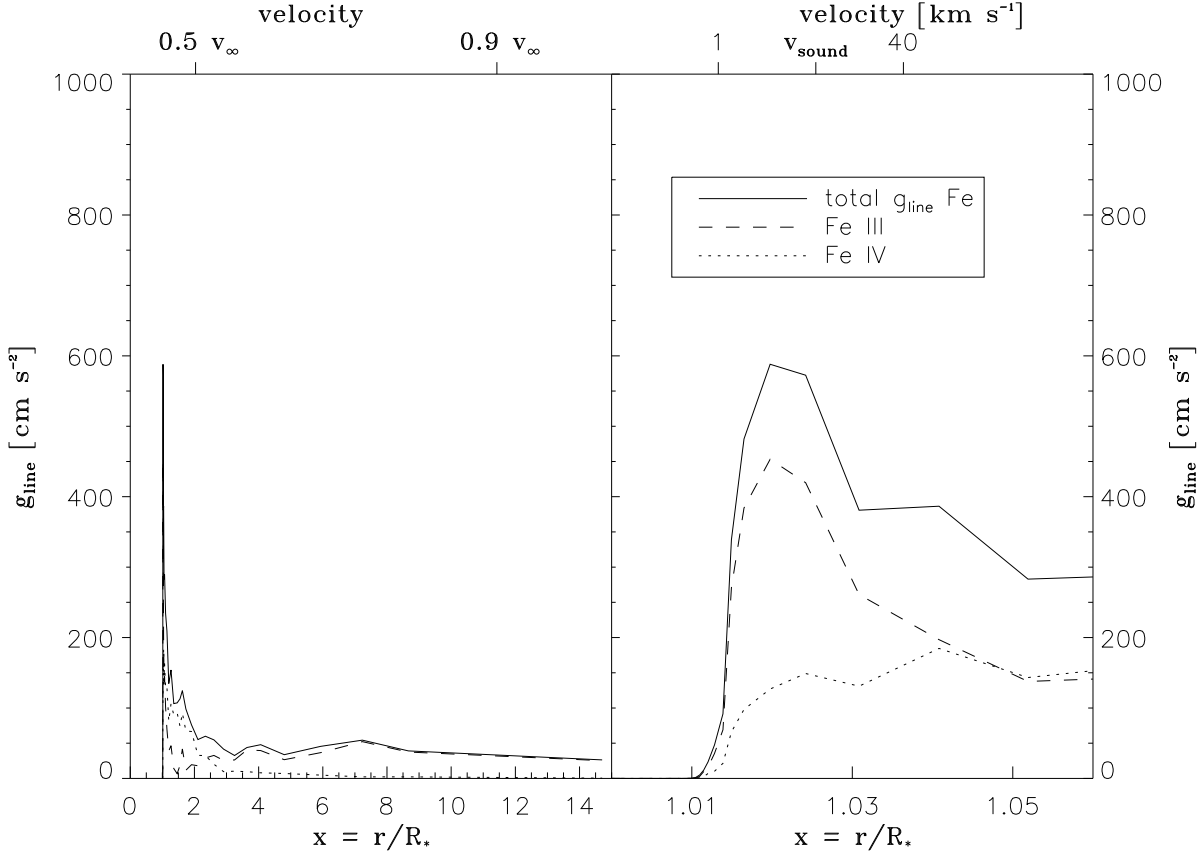}}
 \caption{The contribution of several Fe ions to  
         $g_{\rm L}$ as a function of distance from the stellar surface
         for model A.
          ~{\bf (a)}~The full distance range of $x=1$ to 15.
          ~{\bf (b)}~The region around the sonic point is enlarged.
          The legend indicates the ionization stage.
          Some values for the velocity are indicated on the top of the figure.
          Note that the strongest contribution to $g_{\rm L}$
          below the sonic point is due to Fe~{\sc iii}, although the
          ionization fraction of this ion is less than 10 \%.}
\label{f_AFe}
\end{figure*}

Fig.~\ref{f_ionization} (b) clearly shows that 
Fe~{\sc iv} is the dominant ionization stage in the 
subsonic region of the stellar wind. In the region just 
below the sonic point, the ionization fraction of 
Fe~{\sc iv} is 90~-~100 \%\ whereas that of Fe~{\sc iii} is less
than 10 \%. However, this does not necessarily 
mean that Fe~{\sc iv} is the main line driver. To investigate 
the contribution to the line acceleration $g_{\rm L}$ of the different 
ionization stages of Fe some extra Monte-Carlo simulations 
were performed. One simulation included only the lines of 
Fe~{\sc iii}, another simulation included just the lines of 
Fe~{\sc iv}. The results for $g_{\rm L}$ for Fe~{\sc iii} and 
Fe~{\sc iv} are plotted in Fig.~\ref{f_AFe}.

It is surprising to note that, although Fe~{\sc iv} is 
the dominant ionization stage throughout the wind, most of 
the driving is contributed by Fe~{\sc iii}. Below the sonic 
point Fe~{\sc iii} is clearly the most important iron line driver 
(see Fig.~\ref{f_AFe}(b)).

From the data shown in Figures \ref{f_ionization} and \ref{f_AFe}
we conclude that the mass-loss rate of winds from stars with
$\Teff ~\simeq 27~500$ K is mainly determined by the
radiative acceleration due to Fe~{\sc iii} lines.
This suggests that the bi-stability jump
is mainly due to changes in the ionization balance of Fe.
We test this hypothesis in the next section.


\subsection{The effect of \teff\ on \mdot}
\label{sec:teff}

In the previous section we have shown that the mass loss of
model A is dominated by radiative acceleration due to
Fe~{\sc iii} lines. In this section we investigate changes
in the radiative acceleration due to Fe as $\Teff$ decreases.
This may explain the increase of \Mdot\ near the 
bi-stability jump. 
To this purpose we compare 
the ionization and $g_{\rm L}$ of models A and B in detail.  

The ionization balance of model B is shown in 
Fig.~\ref{f_ionization}(c) and (d). It shows that, due to 
a lower temperature, the decrease of the Fe~{\sc iv} fraction drops 
to smaller values than for model A, which was shown in 
Fig.~\ref{f_ionization}(b). 
The ionization fraction of Fe~{\sc iii} below 
the sonic point in the case of model B is up to almost 40 \%. 
To see whether this extra amount of Fe~{\sc iii} can cause 
the increase in the line acceleration, we must look at $g_{\rm L}$ of Fe 
for model B. 

Since model A and B have different \teff\ at the same $L_*$, 
they have a different radiative surface flux. The radiative
acceleration will be proportional to this flux.
In order to compare the 
values of $g_{\rm L}$ of the two models, we scale the results to a
flux of a $\Teff$ = 25~000 K model. So   

\begin{equation} 
\label{eq:glnorm}
g_{\rm L}^{\rm norm}~=~g_{\rm L}~\left( \frac{25000}{\teff}\right )^4
\end{equation}
Since $\Teff ^4 \propto R_*^{-2}$ for constant luminosity,
this is also a scaling to the Newtonian gravity of the models.

\begin{figure*}
 \centerline{\psfig{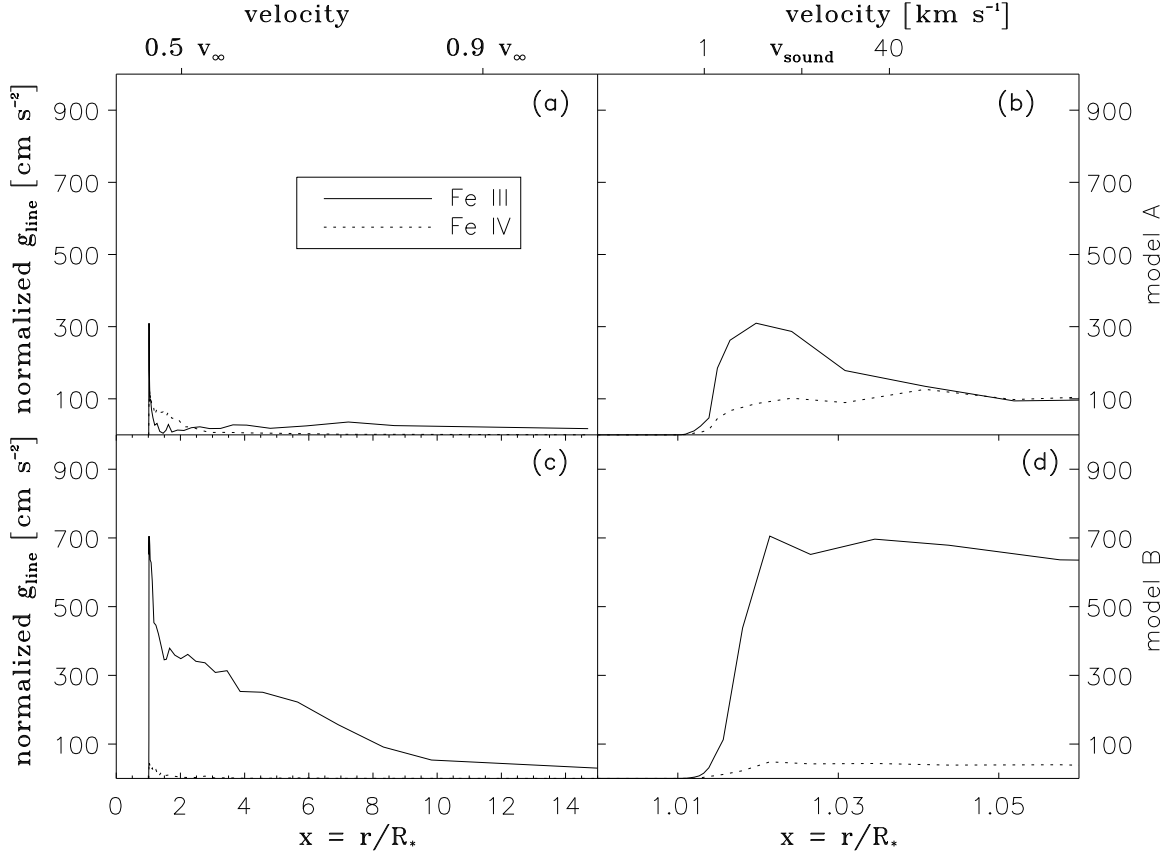}}
 \caption{Normalized $g_{\rm L}$ of Fe as a function of distance from the stellar surface for
          the models A and B.
          ~{\bf (a)}~Normalized $g_{\rm L}$ for the different Fe ionization
          stages of model A. The legend indicates the ionization stage.
          Some values for the velocity are indicated on the top of the figure.
          ~{\bf (b)}~model A, enlarged around the sonic point.
          ~{\bf (c)}~~Normalized $g_{\rm L}$ for the different Fe ionization
          stages of model B.
          ~{\bf (d)}~model B, enlarged around sonic point.}
\label{f_ABFe}
\end{figure*}

Figure \ref{f_ABFe} shows the normalized $g_{\rm L}$ of Fe for the 
models A (top) and B (bottom). The right hand figures show an
enlargement of the region near the sonic point.
It shows that for model B $g_{\rm L}$ of Fe~{\sc iii} around the sonic point  
is more than a factor two
larger than for model A (see Figs.~\ref{f_ABFe}(b) and (d)). This extra
amount of Fe~{\sc iii} in model B causes an increase in the $total$ 
$g_{\rm L}$ in the subsonic part of the wind also, as can be seen in
Fig.~\ref{f_ABCABC}(b). 

{\it We conclude that the increase in mass loss from model A to B
is due to the larger radiative acceleration (compared to the gravity)
of model B by a larger ionization fraction of Fe~{\sc iii} 
below the sonic point.}


\subsection{The effect of $\vinf$}
\label{sec:vinf}

\begin{figure*}
 \centerline{\psfig{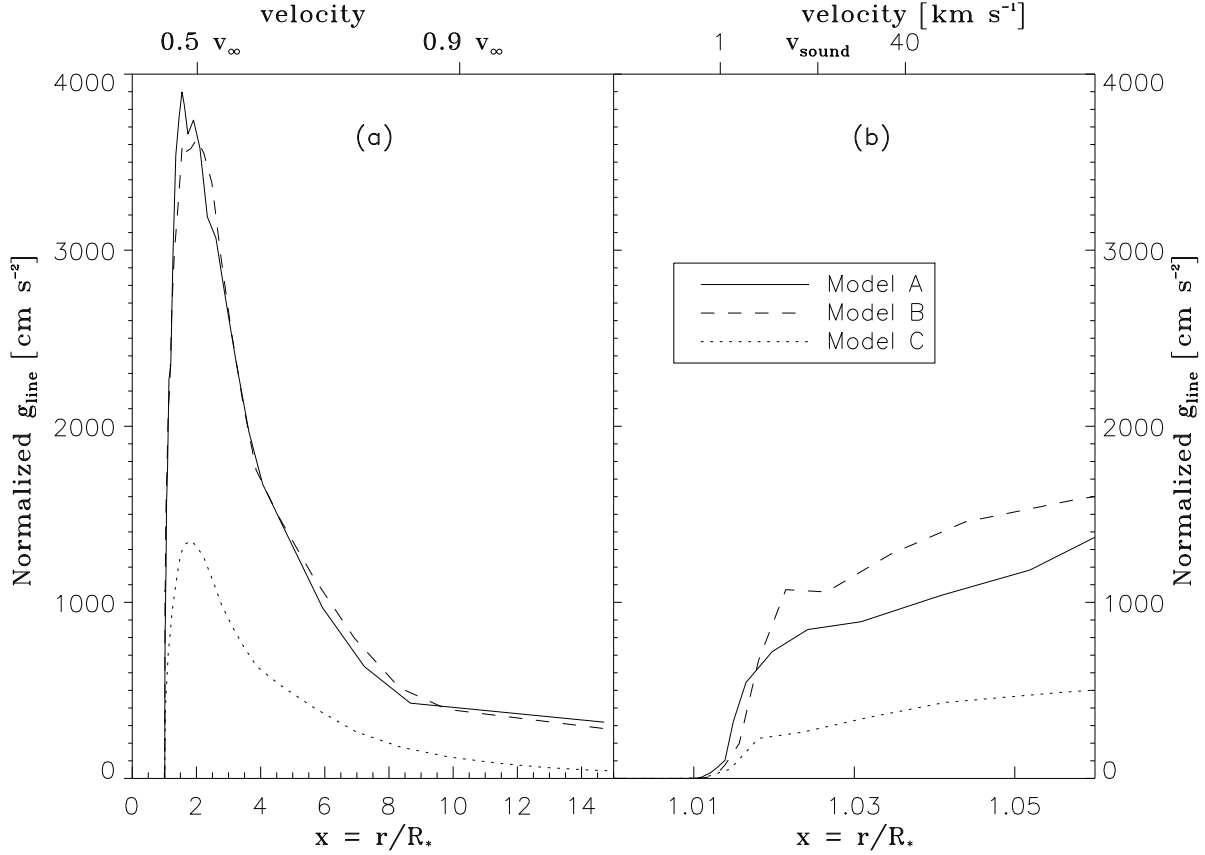}}
 \caption{{\bf (a)}~ The normalized {\it total} $g_{\rm L}$ for the models A, B and C
     as a function of distance. Notice the much smaller radiative acceleration in 
     the supersonic region of model C
     compared to models A and B. 
     {\bf (b)}~An enlargement of the region around the sonic point.
     The sonic point is located around $x = 1.025~r/R_{*}$.
     Notice also the much smaller radiative acceleration in 
     the subsonic region of model C compared to models A and B. This
     is due to the smaller value of $\ratio$ for model C. 
     }
\label{f_ABCABC}
\end{figure*}

Now the effect of $g_{\rm L}$ on $\vinf$ will be examined. Therefore, Model B
is compared to model C.
We remind that models B and C have the same \Teff,
and hence the same radiative flux and gravity,  but 
model C has a twice as small value of $\ratio$ as model B. 
Figure \ref{f_ABCABC}(a) shows the normalized $g_{\rm L}$ for  
models A, B and C. As expected, $g_{\rm L}(r)$ for model C is significantly 
smaller than 
$g_{\rm L}(r)$ for models A and B. This is obviously due to the 
smaller value of \vinf.
The integral $\int g_{\rm L}(r)~dr$ in Fig.~\ref{f_ABCABC}(a) for model A and B
is larger than for model C. The values of $\int g_{\rm L}(r)~dr$ for the models 
are 2.34 $\times$ 10$^{16}$ and 1.92 $\times$ 10$^{16}$ cm$^2$ s$^{-2}$ for models A and B respectively,
and 6.12 $\times$ 10$^{15}$ cm$^2$ s$^{-2}$ for model C. Using Eq. \ref{eq:vinfty} 
and the values of \vesc\ from
column (4) in Table \ref{t:parameters}, the output values for \vinf\ can be obtained from the 
values of the integral of $g_{\rm L}$. 
The derived output values for \vinf\ for the models are 
\vinf\ = 2050, 1860 and 920 km~s$^{-1}$ respectively for the models A,B and C. 
These values are equal within 10 \% to the input values for \vinf\, which 
were indicated in column (5) of Table \ref{t:parameters}.
We can conclude that a smaller value for \vinf\ is indeed consistent
with a smaller value of the integral $\int g_{\rm L}(r)~dr$. 
However, this is not an independent check, since
the calculated line acceleration of optically thick lines 
(in the Sobolev approximation)
is inversely proportional to the Sobolev optical depth 
which is proportional to $(dv/dr)^{-1}$. 
Hence, {\it assuming} a smaller terminal velocity will
automatically result in a smaller calculated line acceleration.

\begin{table*}
 \caption[]{Force multipliers and consistent models. \\
  log $L/\lsun = 5.0$, $M = 20 \msun$, 
  $\Gamma_{e} = 0.130$, $M_{\rm eff} = 17.4~\msun$,  \( \beta = 1 \), solar metallicity.}
  \label{t:forcem}
  \begin{tabular}{cccccccccc}
  \hline
  \noalign{\smallskip}
  $\teff$  & $(\ratio)_{0}$ & $(\ratio)_{1}$ & $(\ratio)_{2}$ & $(\ratio)_{3}$ & $\alpha^{\rm MC}$ & $k^{\rm MC}$ & $(\ratio)_{4}$ & log~$\mdot_{\rm_{CAK}}$ & log~$\mdot_{\rm_{MC}}$\\
  \noalign{\smallskip}
  (K)   & & & & & & & & ($\msun$/yr) & ($\msun$/yr)\\
  \hline

   17 500   &  2.0  & 1.5 & 1.3 & 1.2 & 0.58 & 0.2065 & 1.2 & -6.21 & -6.27\\  
   30 000   &  2.0  & 2.5 & 2.4 & 2.4 & 0.85 & 0.0076 & 2.4 & -6.86 & -6.90\\  
 
 \noalign{\smallskip}
 \hline

  \end{tabular}
\end{table*}

\subsection{A self-consistent solution of the momentum equation}

In earlier sections we have demonstrated that the mass loss
around the bi-stability jump increases. 
As we have used {\it observed} values for
the ratio $\ratio$ in our model calculations, we have not yet
provided a self-consistent explanation of the observed 
bi-stability jump in $\ratio$. 
As a consistency test of our calculations and an attempt to explain the observed
jump in the ratio $\ratio$, we proceeded to solve
the momentum equation of line driven wind models around the 
bi-stability jump. The approach we take is to combine
predicted force multiplier parameters $k$ and $\alpha$ (see below)
from the Monte Carlo calculation with the analytical
solution of line driven winds from CAK.

We calculated the line acceleration $g_{\rm L}$ for several models
with different $\teff$ using the Monte Carlo method.
The values of $g_{\rm L}$ were expressed in terms of the 
force multiplier $M(t)$ (Eq.~\ref{eq:CAK}).
Following CAK we tried to express $M(t)$ in terms
of a power-law fit of the optical depth parameter $t$ (Eq.~\ref{eq:tCAK}).
We found that in the range 20~000 $\le \teff \le$ 27~500,
$M(t)$ is not accurately fit by a power-law, since the ionization
changes over this critical range in $\teff$.
Fortunately, {\it just outside} 
this temperature region, $M(t)$ {\it can} be 
accurately represented in terms of $k$ and $\alpha$, i.e.

\begin{equation}
M^{\rm MC}(t)~=~ k^{\rm MC}~t^{-\alpha^{\rm MC}}
\label{eq:CAKfit}
\end{equation}
Therefore, we have calculated models
with effective temperatures just below ($\teff$ = 17 500 K) and just 
above ($\teff$ = 30 000 K) this critical temperature range.
Self-consistent values of $\vinf$ and $\mdot$ were thus found
in the following way:

%
%

\begin{enumerate}

\item{} We started with an assumed ratio of $\ratio$ = 2.0 (See column (2) in 
Table \ref{t:forcem}).

\item{} The force multipliers $M^{\rm MC}(t)$ were calculated and 
a power-law fit of the type Eq.~\ref{eq:CAKfit} was derived. 
The fit was found to be excellent in the important part of the wind 
between the sonic point and  $v \simeq 0.5 \vinf$. This yielded values
of $\alpha^{\rm MC}$ and $k^{\rm MC}$.
Next, the mass loss and terminal velocity were simultaneously 
calculated from these $\alpha^{\rm MC}$ and $k^{\rm MC}$ parameters using the CAK 
solution of the momentum equation. Note that the solution with the finite disk 
correction (Pauldrach et al. 1986) was {\it not} 
applied, since this is already properly taken into account in the values 
of $\alpha^{\rm MC}$ and $k^{\rm MC}$ calculated in the Monte 
Carlo technique (see Sect.~\ref{sec:method}). 
The superscript, MC, to the force multiplier parameters was added to 
avoid confusion with $k$ and $\alpha$ for a point-like source used
by e.g. Kudritzki et al. (1989).  
The ratio $\ratio$ can be derived from the simple CAK formulation:

\begin{equation}
\frac{\vinf}{\vesc}~=~\sqrt{\frac{\alpha^{\rm MC}}{1~-~\alpha^{\rm MC}}}
\end{equation}
The value for $\alpha^{\rm MC}$ for the model of 30,000 K is significantly higher
than values for $\alpha$ that were calculated before (e.g. Pauldrach et al. 1986), since 
the finite disk is already included in the $\alpha^{MC}$-parameter!

\item{} The new calculated terminal velocity ratio $\ratio$ 
(column (3) of Table \ref{t:forcem}) was used in the next iteration.   

\item{} New mass-loss rates were calculated from the MC approach using the procedure 
as explained in Sect.\ref{sec:determination}. The mass-loss rates are equal within 15 \% 
to the mass-loss rates that can be calculated using the expression for \mdot\ of CAK 
using $\alpha^{\rm MC}$ and $k^{\rm MC}$ .

\item{} The above procedure (step 1. through 4.) was repeated until 
convergence was reached. After four iterations, the 
ratio $\ratio$ did not change anymore. The intermediate values of $\ratio$ are given in
columns (3), (4) and (5) of Table \ref{t:forcem}. The final value 
for the ratio $\ratio$ is given in column (8). For the hot
model (\teff\ = 30 000 K) the final ratio $\ratio$ equals 2.4; for the cool 
model (\teff\ = 17 500 K) $\ratio$ = 1.2. These values are within 10 \% 
of the observed values of $\ratio$, i.e. 2.6 and 1.3 respectively. 

\item{} CAK mass-loss rates were also calculated from the resulting {\it final}
force multiplier parameters $k^{\rm MC}$ and $\alpha^{\rm MC}$ (given in columns (6) and (7)
of Table \ref{t:forcem} and the {\it final} mass-loss rates are
given in column (9) of this Table. Note that the values of
\mdot\ are only marginally different from the mass-loss rates that were calculated from
the Monte Carlo approach (column (10) of Table~\ref{t:forcem}).

\end{enumerate}

In summary; we have self-consistently calculated values of
$\vinf$ and $\mdot$ of two models located at either side of the bi-stability jump.
We have found a jump in 
terminal velocity $\ratio$ of a factor of two, similar as 
observed by Lamers et al. (1995). Moreover, the mass-loss rates
calculated from the CAK formulation
are consistent with those obtained from our Monte Carlo approach.
{\it This implies that the origin of the observed change in the ratio $\ratio$ of
a factor of two around spectral type B1 is identical to the predicted jump
in mass-loss rate of a factor of five due to the recombination of
Fe~{\sc iv} to Fe~{\sc iii}.}


\subsection{Conclusion about the origin of the bi-stability jump}

From the results and figures presented above we conclude 
that the mass-loss rate of early-B supergiants near the 
bi-stability jump is mainly determined by the radiative
acceleration by iron. Although Fe~{\sc iv} is the dominant
ionization stage in the atmosphere of stars near 25~000 K, it is
Fe~{\sc iii} that gives the largest contribution to the 
subsonic line acceleration. This is due to the number of
effective scattering lines and their distribution in
wavelengths, compared to the energy distribution from the
photosphere. This implies that the mass-loss rates of
B-supergiants are very sensitive to the ionization 
equilibrium of Fe in the upper photosphere. 
Our models show that the ionization fraction of Fe~{\sc iii}
increases drastically between \Teff = 27~500 and 25~000 K.
This causes an increase in the line acceleration below the
sonic point and in turn increases the mass loss near the bi-stability jump.




\section{Bi-stability and the variability of LBV stars}
\label{sec:lbv}

Luminous Blue Variables (Conti 1984) are massive stars 
undergoing a brief, but important stage of 
evolution. During this period they suffer severe mass 
loss with \mdot\ values of up to $10^{-4} \msunyr$. 
LBVs are characterized by typical variations in the order of 
$\Delta V$ of 1 to 2 magnitudes. Nevertheless, the total bolometric
luminosity of the star $L_*$ seems to be about constant. The 
reason for the typical LBV variations is still unknown. For reviews see
Nota \& Lamers (1997).

Leitherer et al. (1989) and de Koter et al. (1996) 
have shown that it must be the actual radius of 
the star that increases during these typical variations. 
Therefore, $\teff$ decreases during the variations, if $L_*$ is about
constant. In this paper, we have calculated the mass-loss behaviour 
for normal OB supergiants as a function of \teff. Despite many 
differences between OB supergiants and LBVs, we can retrieve valuable
information about the behaviour of \mdot\ during a typical LBV 
variation by investigating the \mdot\ behaviour of normal OB 
supergiants, since both types of stars are located in the same 
part of the HRD. Our calculations can be used as a tool to 
understand the mass loss changes of an LBV in terms of changes 
in \teff\ during such a typical variation (see also Leitherer et al. 1989).

Observations of LBVs show that for some LBVs that undergo 
typical variations \mdot\ is increasing from visual minimum 
to maximum, while for others it is the other way around: \mdot\ 
is decreasing. This ``unpredictable'' behaviour of \mdot\ during 
an LBV variation is not a complete surprise, if one considers our 
\mdot\ values as a function of $\teff$. We have found that in 
the ranges \(\teff = 40~000-30~000~\rm K\) and 
\( \teff = 20~000-12~500~\rm K\), \mdot\ decreases for a 
decreasing \teff, whereas in the interval between 
\(\teff = 30~000-20~000~\rm K\), \mdot\ increases for a decreasing 
\teff. This shows that whether one expects an increasing or  
decreasing \mdot\ during an LBV variation depends on the specific 
range in \teff\ between visual minimum and maximum.
This was already suggested by Lamers (1997), albeit a 
constant value of $\eta$ was anticipated.

Our present calculations cannot be used to model the observed 
LBV variations, because we have assumed solar metallicities, whereas the LBVs
are known to have an enhanced He and N abundance (e.g. Smith et al. 1994).
Moreover, since most LBVs have already suffered severe mass loss
in the past, their $L_*/M_*$ ratio  
will be higher than for normal OB supergiants. This means that LBVs 
are closer to their Eddington limit, which one may expect to have an 
effect on \mdot\ also. These combined effects explain the lack of a 
consistent behaviour of \mdot\ for LBV variations so far. Especially 
since it is not sure that $L_*$ really remains constant 
during the variations (see Lamers 1995).


\section{Summary, Discussion, Conclusions \& Future work}
\label{sec:concl}

We have investigated the nature of the observed jump in the 
ratio $\ratio$ of the winds of supergiants near spectral type B1.


Calculations for wind models of OB supergiants show that around 
$\teff = 25~000~\rm K$ the mass-loss rate \mdot\ jumps due to an 
increase in the line acceleration of Fe~{\sc iii} below the sonic point. 
This jump in \mdot\ is found in three different series of models. 
In all cases, the wind efficiency number $\eta= \Mdot \vinf /(L_*/c)$ increases 
significantly, by about a factor of 2~to 3, if \Teff\ decreases from
about 27~500 K to about 22~500 K. 
Observations show that the ratio 
$\ratio$ drops by a factor of two  around spectral type B1. Applying 
these values for $\ratio$, we predict a bi-stability jump in 
\mdot\ of about a factor of five. So \Mdot\ is expected to
increase by about this factor between 27~500 and 22~500 K.

We have argued that the mass loss is determined by the 
radiative acceleration in the subsonic part of the wind, i.e. below
$r \simeq 1.03 R_*$. We found that this radiative acceleration
is dominated by the contribution of the Fe~{\sc iii} lines. Therefore
\mdot\ is very sensitive to both the metal abundance
and to the ionization equilibrium of Fe. Our models show that
the ionization fraction of Fe~{\sc iii} and the subsonic
radiative acceleration increases steeply between \Teff = 27~500 and
25~000 K. This explains the calculated increase in \Mdot\ in this
narrow temperature range.

The exact temperature of the bi-stability jump is somewhat ambiguous. 
Observations indicate that the jump occurs around spectral type B1, 
corresponding to $\teff$~$\simeq$~21~000~K (Lamers et al. 1995). 
If one would not completely
trust the value of $\ratio$ for the star HD 109867 (number 91 in Lamers et al. (1995)), 
because 
of its relatively large error bar, then $\teff$ of the observed 
jump can easily occur at
a few kK higher. In fact we cannot expect the bi-stability jump to occur
at one and the same temperature for all luminosity classes, because the
jump is sensitive to the ionization balance (mainly of Fe~{\sc iii}) in the
subsonic region of the wind and hence to the gravity of the star. 
Our models predict that the jump will occur near $\Teff$ $\simeq$~25~000~K. 
However, this is sensitive to the assumptions of the models:
the adopted masses and luminosities and to
the assumption of the modified nebular 
approximation for the calculation of the ionization equilibrium of iron
(see Sect.~\ref{sec:isa}). 

A more consistent
treatment of the ionization and excitation equilibrium of
the Fe-group elements may have two effects: 
{\em  i)} \mdot\ predicted from $\Delta L$ may alter, and
{\em ii)} $\teff$ at which the ionization ratio of e.g. 
          Fe~{\sc iii}/{\sc iv} flips, may shift.
Nevertheless, in view of the very encouraging results using the 
modified nebular approximation in the modeling of UV metal-line 
forests (de Koter et al. 1998), we expect the error in \teff\ 
at which the dominant ionization of Fe switches from {\sc iv} 
to {\sc iii} to be at most a few kK. Furthermore, if a more 
consistent treatment would yield a change in \mdot\, this 
would most likely produce a systematic shift. Since we are 
essentially interested in relative shifts in $\mdot$, we do not 
expect that our conclusions regarding the nature of the bi-stability 
jump would be affected.

It is relevant to mention that Leitherer et al. (1989) calculated 
atmospheric models for LBVs and suggested that the recombination 
of iron group elements from doubly to singly ionized stages, which 
according to them, occurs around \teff\ = 10~000 K, can explain
\mdot\ increases when LBVs approach their maximum states. We
have found a Fe~{\sc iii}/{\sc iv} ionization/recombination
effect around \teff\ = 25~000 K for normal supergiants. We
also anticipate that somewhere, at a lower value of \teff\, a similar
ionization/recombination effect will occur for Fe~{\sc ii}/{\sc iii},
causing a {\it second} bi-stability jump. 
Lamers et al. (1995) already
mentioned the possible existence of a second bi-stability jump around
\teff\ = 10~000 K from their determinations of \ratio, but
the observational evidence for this second jump was meagre.
Possibly, this second jump {\it is} real and we anticipate that this
second jump could very well originate from a Fe~{\sc ii}/{\sc iii} 
ionization/recombination effect. 

Furthermore, we have shown that the elements C, N and O are important
line drivers in the {\it supersonic} part of the wind, whereas the 
{\it subsonic} part of the wind is dominated by the line acceleration
due to Fe. 
\footnote{
When this study was finished we received a preprint by Puls et al. (1999)
who have independently found that the Fe-group elements control the 
line acceleration in the inner wind part, whereas light ions dominate the
outer part.}
Therefore, we do {\it not} expect CNO-processing to have a large
impact on \mdot, but it might have impact on the terminal velocities.

Finally, we would like to add that our calculations for 
\mdot\ around  $\teff = 25~000~\rm K$ have only 
been performed for {\it one} value of $M_*$, $L_*$ and H/He abundance. 
\mdot\ is expected to depend on these stellar parameters, so calculating 
mass-loss rates for a wider range of stellar parameters 
will provide valuable information on the size of the bi-stability 
jump in \vinf\ and \mdot\ and will allow us to constrain its 
amplitude and location in the HRD.


\begin{acknowledgements}

We thank an anonymous referee for several useful suggestions.
JV acknowledges support from the Netherlands Foundation for Research in
Astronomy (NFRA) with financial aid from the Netherlands Organization for
Scientific Research (NWO).
AdK acknowledges support from a NWO ``Pionier'' grant to L.B.F.M.
Waters; from NWO Spinoza grant 08-0 to E.P.J. van den Heuvel and from
a grant from the NASA Astrophysics Data Program NRA 95-ADP-09 to S.R. Heap.

\end{acknowledgements}


\begin{thebibliography}{}


\bibitem[Abbott]{abbott82}
       Abbott D.C., 1982, ApJ 259, 282

\bibitem[Abbott \& Lucy 1985]{abbott85}
       Abbott D.C., Lucy L.B., 1985, ApJ 288, 679

\bibitem[Castor et al. 1975]{castor75}
       Castor J.I., Abbott D.C., Klein R.I., 1975, ApJ 195, 157

\bibitem[Conti 1984]{conti84}
        Conti P.S., 1984, in: Maeder A., Renzini A. (eds.) Proc. IAU
        Symp. 105,  Observational Tests of Stellar Evolution
        Theory, Kluwer, Dordrecht, p. 233 

\bibitem[de Koter et al. 1993]{dekoter93}
       de Koter A., Schmutz W., Lamers H.J.G.L.M., 1993, 
       AAP 277, 561

\bibitem[de Koter et al. 1996]{dekoter96}
       de Koter A., Lamers H.J.G.L.M., Schmutz W., 1996, A\&A 306, 501

\bibitem[de Koter et al. 1997]{dekoter97}
       de Koter A., Heap S.R., Hubeny I., 1997,
       ApJ 477, 792

\bibitem[de Koter et al. 1998]{dekoter98}
       de Koter A., Heap S.R., Hubeny I., 1998,
       ApJ 509, 879

\bibitem[Drew 1989]{drew89}
        Drew J.E., 1989, ApJS 71, 267

\bibitem[Groenewegen \& Lamers]{groen89}
        Groenewegen M.A.T., Lamers H.J.G.L.M., 1989, A\&AS 79, 359

\bibitem[Kudritzki et al. 1989]{kudritzki89}
       Kudritzki R.-P., Pauldrach A.W.A., Puls J., Abbott D.C., 1989, 
       AAP 219, 205

\bibitem[Kurucz 1988]{kurucz88}
        Kurucz R.L., 1988, IAU Trans., 20b, 168

\bibitem[Lamers 1995]{lamers95}
        Lamers H.J.G.L.M., 1995, in: Astropysical Applications of Stellar Pulsations,
        ASP Conf.Ser. 83, 176 

\bibitem[Lamers 1997]{lamers97}
        Lamers H.J.G.L.M., 1997, in: Luminous Blue Variables: Massive Stars in Transistion, 
        ASP Conf.Ser. 120, 76

\bibitem[Lamers \& Cassinelli]{lamcas99}
        Lamers H.J.G.L.M., Cassinelli J.P., 1999, 
        in: Introduction to Stellar Winds, Cambridge Univ. Press, Chapter 3

\bibitem[Lamers \& Pauldrach]{lamers91}
        Lamers H.J.G.L.M., Pauldrach A.W.A., 1991, A\&A 244, L5

\bibitem[Lamers et al. 1995]{lamers95}
        Lamers H.J.G.L.M., Snow T.P., Lindholm D.M., 1995, ApJ 455, 269

\bibitem[Lamers et al. 1999]{lamers99}
        Lamers H.J.G.L.M., Vink J.S., de Koter A., Cassinelli J.P., 1999, 
        in: Variable and Non-Sperical Stellar Winds in Luminous Hot Stars, 159

\bibitem[Leitherer et al. 1989]{leith89}
        Leitherer C., Schmutz W., Abbott D.C., Hamann W.R., Wessolowski U., 1989, ApJ 346, 919

\bibitem[Lucy 1998]{lucy98}
        Lucy L.B., 1998, in: Cyclical Variability in Stellar Winds, ESO ASS Proc 22, 16

\bibitem[Lucy \& Abbott 1993]{lucy93}
        Lucy L.B., Abbott D.C., 1993, ApJ 405, 738

\bibitem[Mazzali \& Lucy 1993]{mazz93}
        Mazzali P.A., Lucy L.B., 1993, A\&A 279, 447

\bibitem[Nota \& Lamers]{nota97}
        Nota A., Lamers H.J.G.L.M., 1997, Luminous Blue Variables: Massive Stars in Transition,
        ASP Conf.Ser. 83.

\bibitem[Pauldrach \& Puls]{paul90}
        Pauldrach A.W.A., Puls J., 1990, A\&A 237, 409

\bibitem[Pauldrach et al.]{paul86}
        Pauldrach A.W.A., Puls J., Kudritzki, R.P., 1986, A\&A 164, 86

\bibitem[Puls et al. 1996]{puls96}
        Puls J., Kudritzki R.P., Herrero A., et al., 1996, A\&A 305, 171

\bibitem[Puls et al.]{puls99}
        Puls J., Springmann U., Lennon, M, A\&A submitted

\bibitem[Schmutz 1991]{schm91}
        Schmutz W., 1991. In: Stellar Atmospheres: Beyond Classical Models,
        eds. Crivellari L., Hubeny I., Hummer D.G., NATO ASI Series C, Vol. 341, 191

\bibitem[Schmutz et al. 1990]{schm90}
        Schmutz W., Abbott D.C., Russell R.S., Hamann W.-R., Wessolowski U., 1990,
        ApJ 355, 255

\bibitem[Smith et al. 1994]{smith94}
        Smith L.J., Crowther P.A., Prinja R.K., 1994, A\&A 281, 833

\end{thebibliography}
\end{document}